\documentclass[sigconf]{acmart}
\usepackage{appendix}
\usepackage{multirow}

\usepackage[utf8]{inputenc}
\usepackage[normalem]{ulem}
\usepackage{subcaption} 
\usepackage{float} 
\usepackage{xcolor}
\usepackage{tikz}
\definecolor{dot1}{HTML}{7bff59}
\definecolor{dot2}{HTML}{DAB1DA}
\definecolor{dot3}{HTML}{ffb654}

\usepackage{framed} 
\usepackage{enumitem}

\definecolor{promptbg}{gray}{0.92}
\definecolor{promptrule}{gray}{0.55}

\newenvironment{promptbox}{%
  \MakeFramed{\advance\hsize-\width \FrameRestore}%
  \ttfamily\small\noindent\ignorespaces
}{%
  \endMakeFramed
}

\begin{document}


\copyrightyear{2026}
\acmYear{2026}
\setcopyright{cc}
\setcctype{by}
\acmConference[HCOMP 2026]{2026 ACM Conference on Human-AI Complementarity and Alignment}{September 27--30, 2026}{Alexandria, VA, USA}
\acmBooktitle{2026 ACM Conference on Human-AI Complementarity and Alignment (HCOMP 2026), September 27--30, 2026, Alexandria, VA, USA}
\acmDOI{10.1145/3834580.3838737}
\acmISBN{979-8-4007-2894-5/2026/09}


\author{Tao Long}
\authornote{Both authors contributed equally to this research.}
\email{long@cs.columbia.edu}
\affiliation{%
  \institution{Columbia University}
  \city{New York}
  \state{NY}
  \country{USA}
}

\author{Xuanming Zhang}
\authornotemark[1]
\email{billyzhang@cs.columbia.edu}
\affiliation{%
  \institution{Columbia University}
  \city{New York}
  \state{NY}
  \country{USA}
}

\author{Sitong Wang}
\email{sitong@cs.columbia.edu}
\affiliation{%
  \institution{Columbia University}
  \city{New York}
  \state{NY}
  \country{USA}
}

\author{Zhou Yu}
\email{zy2461@columbia.edu}
\affiliation{%
  \institution{Columbia University}
  \city{New York}
  \state{NY}
  \country{USA}
}

\author{Lydia B. Chilton}
\email{chilton@cs.columbia.edu}
\affiliation{%
  \institution{Columbia University}
  \city{New York}
  \state{NY}
  \country{USA}
}

\title{DoubleAgents: Human-Agent Alignment\\ in a Socially Embedded Workflow
}


\renewcommand{\shortauthors}{Long and Zhang et al.}

\begin{abstract}

Aligning agentic AI with user intent is critical for delegating complex, socially embedded tasks, yet user preferences are often implicit, evolving, and difficult to specify upfront. We present DoubleAgents, a system for human–agent alignment in coordination tasks, grounded in distributed cognition. DoubleAgents integrates three components: (1) a coordination agent that maintains state and proposes plans and actions, (2) a dashboard visualization that makes the agent’s reasoning legible for user evaluation, and (3) a policy module that transforms user edits into reusable alignment artifacts, including coordination policies, email templates, and stop hooks, which improve system behavior over time. We evaluate DoubleAgents through a two-day in-lab interactive simulation study (n=10), three real-world deployments, and a technical evaluation. Participants' comfort in offloading tasks and reliance on DoubleAgents both increased over time, correlating with the three distributed cognition components. Participants still required control at points of uncertainty —  edge-case flagging and context-dependent actions. We contribute a distributed cognition approach to human-agent alignment in socially embedded tasks. We further introduce interactive simulation as a methodological testbed for rapid iteration and alignment testing of agentic systems.
\end{abstract}





\begin{teaserfigure}
  \includegraphics[width=1\textwidth]{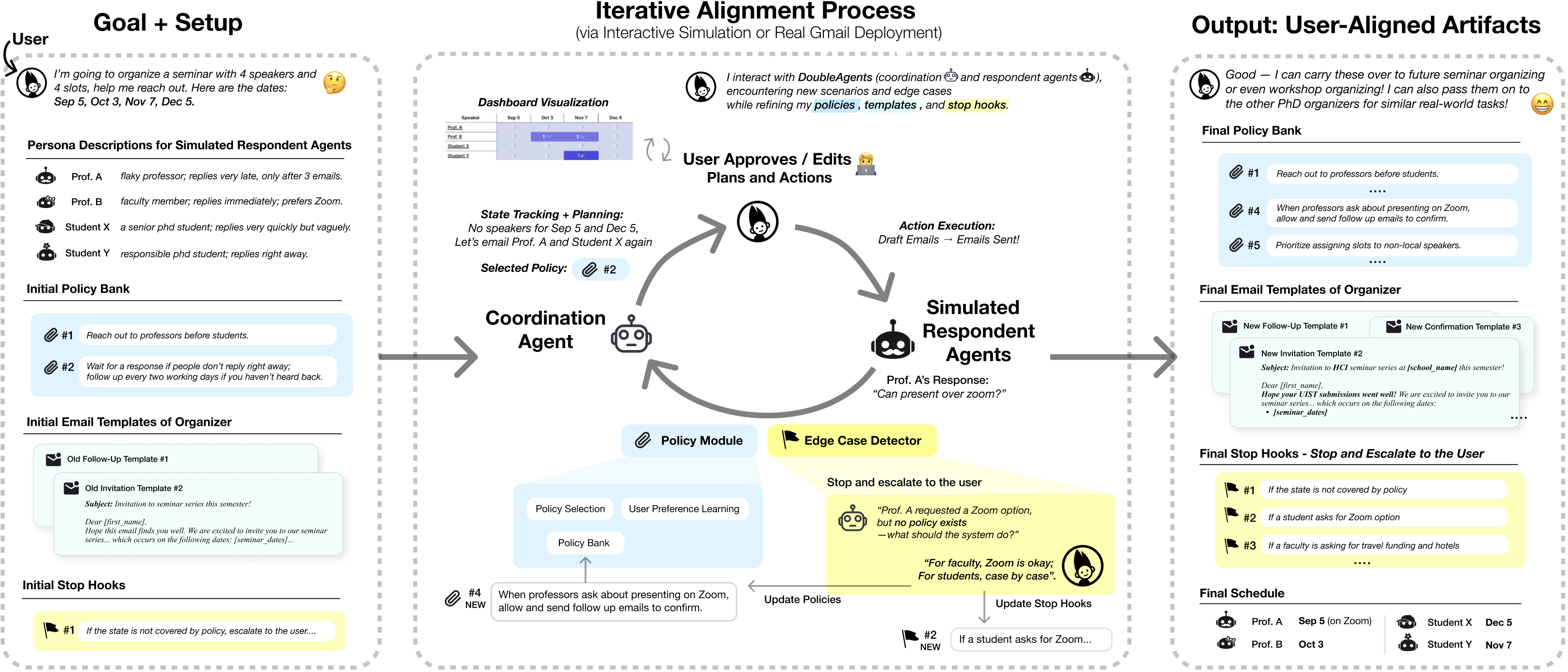}
  \caption{DoubleAgents aligns agentic AI with user preferences for socially embedded coordination tasks such as organizing seminars. From an initial setup (left), including a goal, simulated personas, and seed policies and templates, users iteratively align the system (middle) by reviewing proposed plans, approving actions, handling edge cases, and refining policies, templates, and stop hooks across repeated runs. This produces (right) both a finalized speaker-slot allocation and a set of user-aligned artifacts — policies, email templates, and stop hooks — that persist and transfer to future coordination tasks.}
  \vspace{5px}
  \label{fig:teaser}
\end{teaserfigure}

\maketitle

\section{Introduction}
Alignment between humans and AI is the foundation of useful agentic systems. To act on a user's behalf, an agent must understand intent that is often implicit, evolving, and context-dependent — internalizing tacit norms of when (not) to act, when (not) to ask, and how to match a user's style without being told each time. Without alignment, small misunderstandings scale into systemic problems across workflows and social relationships, eroding the trust that makes delegation possible.

Even well-aligned AI still needs human oversight, both to correct errors and to guide new edge cases in complex, real-world settings. The end goal of alignment is not to fully replace human judgment but to let humans offload as much as possible to AI while maintaining control. How much is offloaded should be up to the human: the more aligned the system, the more confidently a user can delegate. But alignment cannot be specified upfront. Users often do not know what they want until they see the system act, and their preferences evolve as the task unfolds. The central question is  \textit{how humans and AI can work together through repeated interaction and refinement to reach alignment and then comfortable offloading}.

Distributed cognition~\cite{10.1145/353485.353487,hutchins1996distributed} offers a principled framework for designing this kind of human-AI partnership. Research on complex sociotechnical systems has shown that for many hard problems, like flying an airplane, no single individual is fully responsible for the cognition. Instead, cognition is distributed across humans and artifacts: the pilot, co-pilot, autopilot, cockpit instruments, ground control, and maintenance crews all contribute. Three components make this work. First, {\protect\tikz[baseline=-0.5ex] \protect\draw[fill=dot1, draw=black, line width=0.0mm] (0,0) circle (0.1cm);} \textit{tracking state}: the system must maintain a shared sense of where things stand, as cockpit instruments monitor altitude, speed, and position in real time. Second, {\protect\tikz[baseline=-0.5ex] \protect\draw[fill=dot2, draw=black, line width=0.0mm] (0,0) circle (0.1cm);} \textit{representing information}: critical state must be displayed in the right form so actors can make judgments at a glance, as cockpit displays show system overviews rather than raw data. Third, {\protect\tikz[baseline=-0.5ex] \protect\draw[fill=dot3, draw=black, line width=0.0mm] (0,0) circle (0.1cm);} \textit{adapting policies}: all actors follow procedures and checklists that encode accumulated knowledge about safe operation, as aviation checklists evolve after incidents to cover new situations.

We apply this framework to design DoubleAgents, an interactive system for human-agent alignment in socially embedded coordination tasks.
The name reflects the system's two types of AI agents: a {\textit{coordination agent}} that manages the task on behalf of the user, and {\textit{simulated respondent agents}} — either real recipients in live deployment or simulated AI agents that stand in for real people during rehearsal runs. This dual mode lets users iterate on alignment through either live deployment with realistic feedback or simulation that compresses weeks into minutes before real social stakes are involved. 
Our domain is organizing a university speaker seminar series, a task where a human must monitor evolving situations over weeks, follow up with the nonresponsive, and handle unexpected scenarios, all while navigating real social stakes (you cannot unsend an email, nor be too pushy or too passive). 

DoubleAgents distributes cognition through three mechanisms: {\protect\tikz[baseline=-0.5ex] \protect\draw[fill=dot1, draw=black, line width=0.0mm] (0,0) circle (0.1cm);}  \textit{a coordination agent} that tracks all state and proposes next actions, serving as the organizer's memory and reasoning partner; {\protect\tikz[baseline=-0.5ex] \protect\draw[fill=dot2, draw=black, line width=0.0mm] (0,0) circle (0.1cm);}  \textit{a dashboard} with temporal, social, and procedural views that make the agent's reasoning legible so the user can verify and evaluate proposals; and {\protect\tikz[baseline=-0.5ex] \protect\draw[fill=dot3, draw=black, line width=0.0mm] (0,0) circle (0.1cm);}  \textit{a policy module} that turns users' direct edits into reusable alignment artifacts (coordination policies, email templates, and stop hooks) that improve over repeated runs through either live deployment or simulation. 





\section{Related Works}

\subsection{Alignment in Human-AI Systems}

As AI agents take on more complex tasks, ensuring they align with human values has become a central research challenge. Following the Bidirectional Human–AI Alignment framework \cite{bidirectionalalignment}, we view alignment in agentic settings as an ongoing co-adaptation process. On the system side, agents must be continuously shaped through interaction history and curated context so their actions reflect a user's goals, preferences, and social norms \cite{zhang2025towards,gao2024aligning}. On the human side, users must develop appropriate mental models and usage patterns to effectively direct, evaluate, and calibrate delegation \cite{vanderlyn2025understanding,andrews2023role}.
Operationalizing this bidirectional view requires mechanisms that let users specify intent, oversee execution, and revise expectations over time. \citet{terry2023interactive} decomposes this challenge into specification, process, and evaluation alignment, capturing the full arc from goal communication through output verification. Subsequent work has addressed each stage: \citet{shankar2024validates} shows with EvalGen that users' evaluation criteria themselves drift as they observe model outputs, while \citet{epperson2025interactive} supports process-level steering with AGDebugger, enabling what-if probing and checkpoint-based resets to diagnose misalignment in multi-agent workflows. Together, these efforts outline a design space where guardrails, prompt optimization, and interactive debugging serve as complementary post-hoc alignment mechanisms.

Constitutional AI offers a different strategy, replacing constant human feedback with rule-based self-supervision using explicit principles \cite{bai2022constitutional}. This has been expanded into participatory frameworks such as Collective Constitutional AI, where large groups co-author alignment rules to reflect diverse values \cite{huang2024collective}, and ConstitutionMaker, which helps users turn natural language feedback into usable policies \cite{petridis2024constitutionmaker}. DoubleAgents contributes to this space by embedding alignment into an interactive process. Following Suchman's view of plans as situated resources rather than fixed programs \cite{suchman1987plans}, policies and stop hooks in our system function as evolving artifacts that users negotiate in context. Through interactions, users iteratively refine coordination policies, email tone, and stop conditions, treating alignment as a continuous, participatory process shaped by the situations in which it unfolds.

\subsection{Agentic AI: From Task Planning to Socially Embedded Workflows}
Delegating complex tasks to AI agents raises concerns around trust, transparency, and alignment \cite{Afroogh2024Trust}, motivating designs that keep humans in the loop. Exposing agent reasoning with user intervention improves both performance and trust \cite{he2025plan}, yet cognitive research suggests humans rarely plan exhaustively upfront \cite{krieger2009coordinating, teevan2016productivity}, pointing to a need for agents that support incremental, context-sensitive planning with human oversight. Whether agents operate through tool use \cite{toolformer} or direct interface manipulation \cite{anthropic2025computeruse, openai2025operator, manusAI, amazon2025novaact}, the difficulty of robust multi-step planning has been consistently highlighted in recent benchmark works \cite{yao2024tau, zheng2024natural, xie2024travelplanner, trivedi2024appworld}, underscoring that oversight is not merely desirable but necessary.

These challenges intensify when agents operate in socially embedded contexts, where correctness depends on interpersonal norms and situated judgment rather than functional specifications. Benchmarks for social intelligence reveal a wide capability gap: SOTOPIA \cite{zhou2024sotopia} found that LLM agents significantly underperform humans on social commonsense, while CASA \cite{qiu2025casa} showed high norm violation rates in agentic settings, suggesting task-directed focus actively crowds out social sensitivity. 
DoubleAgents targets the gap this literature exposes: the absence of end-to-end systems for multi-turn, socially embedded coordination. Rather than maximizing autonomy, our system helps users iteratively author and refine agent behavior through calibrating delegation boundaries, coordination policies, and communication tone, for tasks where social context and timing are integral to correctness.

\subsection{Human Simulations with LLMs}

Recent work has shown that LLM-based multi-agent simulations can exhibit human-like behaviors in social and coordination settings, forming routines, organizing events, and interacting in ways that resemble human reasoning patterns \cite{park2023generative, horton2023largelanguagemodelssimulated, park2022social}. These simulations have been used to approximate human behavior in domains such as economic decision-making \cite{sreedhar_prosocial_iui_2025, sreedhar_ultimatum_hicss2025, aher2023usinglargelanguagemodels}, judicial reasoning \cite{hamilton2023blindjudgementagentbasedsupreme}, and psychology experiments \cite{hewitt2024predicting}. They have also supported applications such as strategic gameplay \cite{meta2022cicero}, adversarial testing \cite{purpura2025buildingsafegenaiapplications}, and value alignment through simulated interactions \cite{pang2024selfalignment}.

In DoubleAgents, LLM-simulated respondents are used to provide a fast and flexible way to surface a range of coordination scenarios, including potential edge cases, so that users can iteratively refine their delegation strategies. These interactions primarily serve to accelerate the alignment process by making system behavior observable early, while similar workflows can also be carried out during real-world deployment with actual participants.

\section{DoubleAgents System}

\begin{figure*}[!h]
\centering
    \includegraphics[width=.75\linewidth]{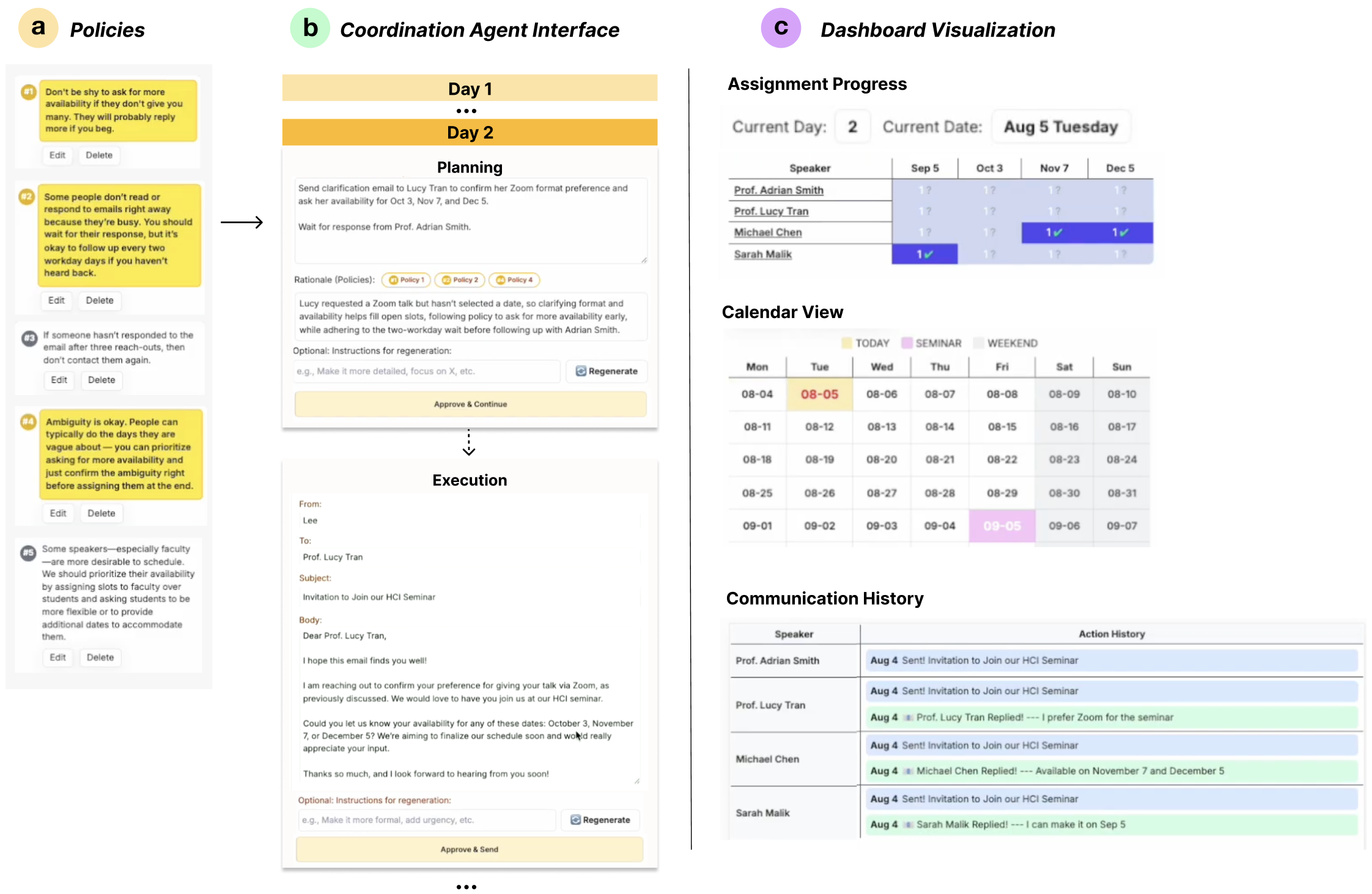}
    \caption{
    A snapshot of DoubleAgents coordinating four speakers across seminar slots. Grounded in \textit{distributed cognition}, the system organizes interaction across three components: (a) {\protect\tikz[baseline=-0.5ex] \protect\draw[fill=dot3, draw=black, line width=0.0mm] (0,0) circle (0.1cm);} Policies - encoding evolving user preferences and procedures; (b) {\protect\tikz[baseline=-0.5ex] \protect\draw[fill=dot1, draw=black, line width=0.0mm] (0,0) circle (0.1cm);} The Coordination Agent - tracking state, plans next steps, suggests actions, and surfaces the policies guiding its decisions; and (c) {\protect\tikz[baseline=-0.5ex] \protect\draw[fill=dot2, draw=black, line width=0.0mm] (0,0) circle (0.1cm);} The Dashboard Visualization - externalizing key information, including assignments, timelines, and communication history, enabling users to quickly interpret the situation. In use, the coordination agent monitors progress, proposes a plan, and explains its rationale; the user reviews the state through the dashboard, verifies that the plan aligns with expectations, and approves it to proceed. Through interactions, policies are refined, allowing the system to better align with the user over time.}
    \vspace{-5px}
    \label{fig:screen_all}
\end{figure*}



DoubleAgents is an interactive system for aligning agentic AI behavior with user preferences in coordination tasks. We ground the system in a concrete use case: organizing a university speaker seminar series (see Appendix~\ref{sec:background} for additional background), a setting where users must manage evolving constraints, follow up with nonresponsive participants, and handle unexpected situations under real social stakes. 
Consistent with our distributed cognition framework, DoubleAgents distributes coordination across three components: a coordination agent that tracks task state and proposes next plans and actions, a dashboard that presents temporal, social, and procedural views to make the agent’s reasoning legible for user evaluation, and a policy module that turns user edits into reusable alignment artifacts (e.g., coordination policies, email templates, and stop hooks). Together, these components support iterative refinement of agent behavior over time, enabling users to progressively align the system through interaction in both simulated and real-world deployment settings.
In this section, we first present a user walkthrough (Section~\ref{subsec:walkthrough}), followed by the system implementation (Section~\ref{subsec:implementation}).

\subsection{User Walkthrough}
\label{subsec:walkthrough}
To illustrate how DoubleAgents operates in practice, we walk through a concrete example. Consider Lee, a PhD student organizing the HCI seminar series at his university. Lee plans to invite four speakers, Prof. Adrian Smith (a famous professor), Prof. Lucy Tran (a rising star professor), Michael Chen (a senior PhD student), and Sarah Malik (a PhD student who is a friend of Lee) to deliver talks during four available time slots: Sep 5, Oct 4, Nov 7, and Dec 5. We demonstrate three example cases of how Lee uses DoubleAgents to solve the problem: (a) starting email bunch, (b) handling edge cases, and (c) following up with non-responsive speakers.

\subsubsection{\textbf{Setting up}}
Lee begins at the landing page (Figure~\ref{fig:landing_page}), where he specifies his goal: ``I am trying to organize a seminar with four speakers and four slots—help me reach out.'' He then provides additional details, including the seminar location and the dates of the four time slots. Next, Lee enters information for each speaker, such as their names and email addresses. When operating in a simulated setting, Lee can optionally include brief persona descriptions to approximate the behavior of the speaker. For example, Prof. Adrian Smith is described as ``a famous professor who is very busy and often does not reply to emails,'' while Sarah Malik is ``a friend who is generally responsive.'' In real-world deployment, these personas are not required, as interactions occur with actual participants. Finally, Lee uploads several past emails he has sent to seminar speakers. This allows the system to adapt to his communication style, generating messages that are more casual for friends and more formal for faculty.

With this information, the system could send emails to speakers on Lee's behalf. But Lee is nervous about this. He wants to make sure the system sends well-written, personalized emails; follows up appropriately if speakers do not respond; nudges people without being pushy; fills all four slots without double-booking; and, critically, does not offend a famous faculty member with an overly casual message. Lee does not yet know how the system will behave, so rather than letting it act in the real world, he starts a simulation that will not send any actual emails. Instead, the simulation lets him see exactly how the system reasons, what plans it proposes, and how it responds to different speaker behaviors, all guided by a set of policies he can inspect and refine. Lee is also shown a list of preset policies in the Policy Panel (Figure~\ref{fig:screen_all}(Left)). These are natural-language rules that will guide the AI's reasoning: for example, when to follow up, how to handle conflicts, and when to give up on a non-responsive speaker. Lee reviews them, finds them reasonable, and is now ready to start the simulation.

\subsubsection{\textbf{DoubleAgents starts and policy selection.}}

When the simulation starts, Lee sees the Control Panel (Figure~\ref{fig:screen_all}(Middle)) alongside a dashboard showing the assignment progress, calendar, and communication history (Figure~\ref{fig:screen_all}(Right)). The calendar shows that in the simulation, today is August 5, and the first seminar date to fill is September 5, so he should start reaching out now. Obviously, nobody has been contacted yet. Lee does not have a clear strategy for who to email first or what to say, so he lets the AI suggest a plan. The coordination agent assesses the current state: no speakers emailed, no availability collected, four slots to fill, and looks for a relevant policy in the existing policy bank, which contains “\textit{If someone hasn’t responded to the email after three reach-outs, then don’t contact them again.}”, “\textit{Don’t be shy to ask for more availability if they don’t give you many. They will probably reply more if you beg.}” and a few others. Then the agent finds one that's most relevant: ``\textit{If we are just beginning and have not yet contacted any speakers for availability, then ask each speaker for their availability across all slots.}'' The selected policy is clearly highlighted to Lee in the Policy Panel (Figure~\ref{fig:screen_all}(Left)), so Lee can see exactly why the AI is about to do what it does. Based on this policy, the AI subsequently proposes a plan (Figure~\ref{fig:plan_gen}): ``Send email to Prof. Adrian Smith, Prof. Lucy Tran, Michael Chen, and Sarah Malik to request availability for slots on Sep 5, Oct 4, Nov 7, and Dec 5.'' 

This makes sense to Lee: scheduling cannot proceed without first collecting availability. So he clicks `Approve \& Continue.' The system then generates four concrete actions, one email per speaker (Figure~\ref{fig:action_gen}), which Lee reviews and approves. Now the AI drafts the actual emails (Figure~\ref{fig:email_gen}). Lee reads the first draft carefully. He notices the date format is inconsistent and wants to add a note mentioning that his advisor recommended the speaker. He makes these edits and clicks `Regenerate.' The revised draft incorporates all his changes. For the next email to Prof. Lucy Tran, the system has already learned Lee's preferred date format and applies it automatically, requiring no further corrections. Lee approves and sends all four emails. Even though these are simulated, Lee is reassured: the AI is learning his style, and he can see exactly what would be sent in the real world.

\subsubsection{\textbf{Handling responses from speakers.}}

In reality, Lee would have to wait days or weeks for responses. In the simulation, time is compressed: a full day in the real world takes only one to two minutes. So Lee quickly sees what happens next. By the end of Day 1, three of the four speakers have responded. Michael Chen and Sarah Malik reply with their availability. Prof. Lucy Tran also replies promptly, but with an unexpected request: ``\textit{Thanks for reaching out, Lee! I was wondering if there might be an option to present via Zoom instead of in person?}'' Prof. Adrian Smith, true to his persona, does not respond at all. The assignment progress tracker immediately updates to reflect who has replied and what slots they are available for. The communication log records every exchange. Lee can see at a glance where things stand and where the gaps are.

\subsubsection{\textbf{Edge cases handling.}}

Prof. Lucy Tran's Zoom request creates a problem: none of the existing policies address remote presentations. The AI does not try to handle this on its own. Instead, the edge case detector flags it and asks Lee for guidance: ``\textit{Prof. Lucy Tran requests a Zoom option. Should I confirm Zoom availability or insist on in-person participation?}'' This is exactly the kind of situation Lee was worried about: the AI encountering something unexpected and making a bad call. But instead of acting autonomously, the system pauses and asks. Lee decides that Zoom is acceptable and types: ``\textit{Zoom is allowed.}'' The AI generates a response to Prof. Lucy Tran, which Lee reviews and sends. Importantly, Lee's decision does not resolve this case alone. The system records it as a stop hook and updates the policy set to include a new rule about allowing Zoom when speakers request it. If a similar situation arises later, for example, Michael Chen also asks about Zoom, the system will already know how to handle it without bothering Lee again. At the end of Day 1, the system generates a daily summary of everything that happened: which emails were sent, who replied, and what decisions were made. Lee reviews it, finds it accurate, and clicks `Continue to Next Day.'

\subsubsection{\textbf{Clarifying, waiting, and following up.}}
The calendar advances. Prof. Adrian Smith has not yet responded, while Prof. Lucy Tran's Zoom details remain unresolved. The AI generates a plan: clarify Zoom logistics with Prof. Tran and continue waiting for Prof. Smith (Figure~\ref{fig:screen_all}(Middle)). Lee checks the rationale, confirms it references the appropriate policies, and approves. The system drafts a follow-up to Prof. Tran, who replies confirming November 7th via Zoom. The dashboard updates accordingly, leaving October 4th as the only open slot.

On Day 3, Prof. Smith has still not responded. The coordination agent, guided by the follow-up policy specifying action after two workdays of silence, drafts a gentle nudge. Lee notices it matches his usual follow-up style: short and to the point, with no unnecessary pressure. This is something the system picked up from uploaded emails and his earlier edits. He approves without changes. Later on, Prof. Smith replies with availability on November 7th, creating a new conflict: three speakers now favor the same date while October 4th remains unfilled. Lee can see the conflict clearly in the assignment progress tracker. From this point onward, the system continues to generate plans grounded in policies, learn from Lee's past choices, and execute actions adaptively, resolving the scheduling conflict and filling the remaining slot until the seminar assignment is complete.

\subsubsection{\textbf{Usage conclusion.}}

By the time the DoubleAgents usage concludes, Lee has not only produced a tentative speaker-to-slot assignment but has also built up a set of refined policies, customized email templates, and executable stop hooks that align with his actual preferences. He now knows how the AI handles follow-ups, edge cases, and conflicts, because he has seen it happen in a safe and simulated environment. When moving to real-world deployment, such as sending actual emails to speakers, these artifacts directly guide the agent’s actions, reducing uncertainty and enabling more reliable coordination. Over time, this process accumulates into a reusable set of alignment artifacts that can be applied and further developed in future coordination tasks.

\subsection{System Implementations}
\label{subsec:implementation}

We implemented DoubleAgents as an interactive simulation system built on the ReAct framework \cite{yao2023react}. The system architecture comprises four key components: (1) a \textit{Simulation Engine} that drives an iterative planning-execution loop through a coordination agent and LLM-based simulated respondent agents, (2) a \textit{Policy Module} that encodes user values as natural language rules, supports policy selection and refinement through stop hooks, and captures implicit user preferences from edit behavior, (3) an \textit{Edge Case Detector} that identifies situations beyond existing policy coverage and escalates them for user clarification, and (4) a \textit{Dashboard Visualization} built on Flask that provides real-time visual feedback on assignment progress, communication history, and scheduling timelines. Users interact with the system via a control panel that supports plan review, action editing, and natural-language queries. Full implementation details, including prompts and system architecture, are provided in Appendix~\ref{sec:appendix_implementation}.

\subsection{Technical Evaluation}
\label{tech-eval}

DoubleAgents enables iterative task planning and action execution through two core mechanisms: (1) \textit{policy selection}, which steers plan generation toward user-aligned coordination strategies; and (2) \textit{edge case detection}, which identifies ambiguous or potentially problematic situations and escalates them for user intervention. We evaluate both at the component level; end-to-end coordination quality is assessed through the user study (Section \ref{sec:user_study}) and deployments (Section \ref{sec:deploy}). Full experimental setups, including input conditions, test-suite construction, models, and prompt templates, are provided in Appendix~\ref{app:tech-eval}. We report the results in this section.

\subsubsection{Policy Selection}
Policy selection retrieves the coordination policies relevant to the current problem state (e.g., unfilled slots, unresponsive speakers) to guide the next planning step. We compare four representations of that state — speaker-only, timeslot-only, their combination, and a summary-based abstraction — on a suite of 30 scenarios drawn from real user-study sessions, scoring F1 against the policies users approved during the study (Appendix~\ref{app:tech-eval}).

As shown in Table~\ref{tab:techeval_policy}, the summary-based representation achieves the highest F1, significantly outperforming all three baselines and indicating that a structured, high-level snapshot of task progress yields more accurate policy retrieval than raw context. Notably, the combined condition does not improve over the simpler single-source variants, suggesting that aggregating low-level context without abstraction is insufficient for effective retrieval. We therefore adopt the summary-based representation as the default input to policy selection; the retrieved policies are passed to the coordination agent to guide plan generation in a policy-aligned manner.

\subsubsection{Edge Case Detection}
Edge case detection determines whether a respondent's message falls outside the scope of existing policies and should be escalated: for example, a Zoom request when no policy addresses remote presentations. We compare three prompting conditions, namely zero-shot, few-shot without policy context, and few-shot with policy context, on a suite of 100 manually validated scenarios, reporting accuracy (Appendix~\ref{app:tech-eval}).

As shown in Table~\ref{tab:techeval_edge}, few-shot prompting with policy context achieves the highest accuracy, significantly outperforming both baselines. Access to explicit policy context lets the model reliably judge whether a scenario is already covered or requires escalation, whereas zero-shot prompting struggles to separate routine variation from true edge cases, and few-shot without policy context tends to over-flag benign messages or miss subtle policy gaps. We adopt the policy-aware few-shot approach as the default, surfacing novel or ambiguous situations for timely user intervention while keeping the coordination workflow smooth.

\begin{table}[!h]
\vspace{-5px}
\begin{minipage}[t]{0.48\columnwidth}
\centering
{\footnotesize
\begin{tabular}{lcc}
\toprule
Prompting Tech. & Mean & SD \\
\midrule
Speaker-only & .27 & .054 \\
TimeSlot-only & .11 & .038 \\
TimeSlot + Speaker & .27 & .017 \\
Summary-based & \textbf{.70}*** & .035 \\
\bottomrule
\end{tabular}
}
\subcaption{Policy Selection (F1 Score).}
\label{tab:techeval_policy}
\end{minipage}
\hfill
\begin{minipage}[t]{0.5\columnwidth}
\centering
{\footnotesize
\begin{tabular}{lcc}
\toprule
Prompting Techni. & Mean & SD \\
\midrule
Zero-shot & .79 & .021 \\ \midrule
Few-shot w/o Policy & .88{\textdagger} & .000 \\
Few-shot w/ Policy & \textbf{.93}** & .015 \\
\bottomrule
\end{tabular}
}
\subcaption{Edge Case Detection (Accuracy)}
\label{tab:techeval_edge}
\end{minipage}
\vspace{-5px}
\caption{Technical evaluation results. Best scores in bold. Significance via independent two-sample t-test: {\textdagger}~$p<0.05$ over one baseline, $^{**}$~$p<0.01$, $^{***}$~$p<0.001$ over all baselines.}
\vspace{-20px}
\label{tab:techeval}
\end{table}
\section{User Study}
\label{sec:user_study}

\subsection{Study Design}
\subsubsection{\textbf{Overview}}
To investigate how users and AI can reach alignment through repeated interaction and refinement --- and how this alignment enables comfortable offloading --- we recruited 10 participants across two sessions over two consecutive days. Each participant completed approximately 2.5 hours total: a 30-minute preparation and setup phase, followed by a 60-minute usage and evaluation session each day. Our goal was to assess whether comfort with delegation evolves over time, how the system's three distributed cognition components support users in learning what is safe to delegate, and where users still need to maintain control. During each session, participants worked in a simulated environment to schedule a seminar with four dates and four speakers, while the system gradually adapted to participants' values and preferences through their policies, email templates, and stop hooks. Participants completed two questionnaires. An \textit{AI perception survey} measured alignment expectations, trust, and comfort with automating planning and execution; it was administered before Session~1, after Session~1, and after Session~2. A \textit{DoubleAgents experience survey} captured UX, task load, and perceptions of system features after each session. Together, these instruments allowed us to track how repeated use shifted alignment, delegation boundaries, and the evolving balance between offloading and control. Detailed study procedures and data analysis methods are provided in Appendix~\ref{sec:userstudyProcedures}.

\begin{figure*}[!h]
    \centering
    \includegraphics[width=1.02\linewidth]{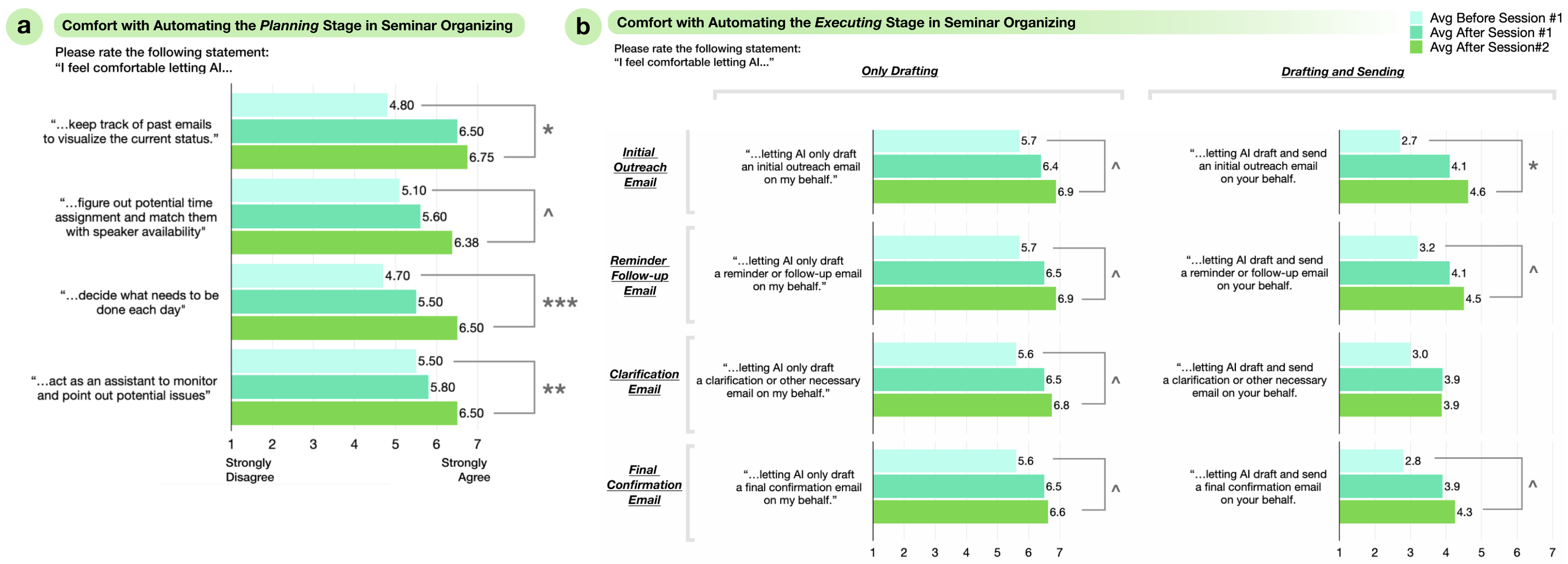}
    \vspace{-18px}
    \caption{Average user-reported scores across 10 participants for comfort with automating \textbf{(a)} \textit{planning} and\textbf{ (b) }\textit{execution} tasks, across sessions over time (1: strongly disagree, 7: strongly agree). $^{\wedge}$~$p < 0.1$, $^{*}$~$p < 0.05$, $^{**}$~$p < 0.01$, $^{***}$~$p < 0.005$.}
    \label{fig:resultUIST}
\end{figure*}
\subsubsection{\textbf{Participants}}
We recruited 10 participants (7 female, 3 male; aged 21--30, $M = 25.7$) from our university community via personal contacts and departmental mailing lists.
All reported daily use of generative AI tools, including AI writing assistants (10/10), coding assistants (7/10), and design tools (5/10). Seven described in-depth knowledge of generative AI models, while three reported a good general understanding of key concepts. Participants' experience with human coordination tasks varied: 2 had moderate to extensive experience, having led team schedules or group planning multiple times; 3 had some experience, having organized small tasks  independently; 4 had minimal experience, primarily helping or shadowing others in coordination tasks; and 1 had no prior experience.

\subsection{Results}

We organize our findings around the central question of how humans and AI reach alignment through repeated interaction and refinement, enabling comfortable offloading. We examine three aspects: whether comfort with delegation evolves over time (\S\ref{sec:comfort}), how the three distributed cognition components support participants in learning what is safe to delegate (\S\ref{sec:learning}), and where participants still need to maintain control (\S\ref{sec:control}).

\subsubsection{\textbf{\uline{Did participants become more comfortable delegating control over time?}}}
\label{sec:comfort}


\paragraph{\textbf{Participants reported increased comfort delegating both planning and execution tasks to AI}}
For planning (Figure~\ref{fig:resultUIST}a), comfort ratings rose across all four sub-tasks (letting the AI track and visualize task state, solve scheduling constraints such as time-slot assignments, reason about context to determine next steps, and monitor for edge cases requiring user attention), with averages climbing from roughly 5/7 before deployment to around 6.5/7 afterward. Constraint solving for slot allocation is a comparatively task-specific capability and showed only marginally significant improvement. In comparison, the planning sub-tasks, tracking, monitoring, and flagging shows strong comfort gains over time, consistent with the distributed-cognition affordances of the coordination agent that help organizers offload ongoing awareness work. 

For execution (Figure~\ref{fig:resultUIST}b), the pattern split by autonomy level: participants were already fairly comfortable letting the system \emph{draft} emails on their behalf, and comfort rose further (from around 5.7/7 to nearly 6.8/7 across all four email types, each at least marginally significant).
By contrast, comfort with the system both \emph{drafting and sending} emails---where the social stakes of an error are higher---started considerably lower and remained so, though it still improved meaningfully (from roughly 3/7 to 4.3/7 across email types). Among the four email types, initial outreach showed strong gains, likely because its content is relatively formulaic and low-variance; follow-up and confirmation emails improved modestly, while clarification emails showed no significant change over the alignment process---an exception we return to in Section 4.2.3.


Participants' perceived reliance on DoubleAgents, as measured by the \textit{DoubleAgents Experience survey}, increased significantly (4.90/7 to 5.75/7, $p=0.02^{*}$). Participants characterized this not as unconditional reliance, but as increased comfort with delegation as the system's behavior became more predictable and legible — feeling both aligned with how they wanted to work and comfortable letting the system act on their behalf. P2 articulated this arc explicitly: \textit{``I would intervene more at first but if it can demonstrate that it's consistently able to do it correctly, I'm going to let it act on its own on my behalf more autonomously.''}

\subsubsection{\textbf{\uline{How did participants use and rate the three distributed cognition components?}}}
\label{sec:learning}
\paragraph{\textbf{{\protect\tikz[baseline=-0.5ex] \protect\draw[fill=dot3, draw=black, line width=0.0mm] (0,0) circle (0.1cm);} Policies and templates helped participants encode their values, reducing effort over time.}}
Carrying over user-defined coordination policies strengthened alignment in planning: participants found the AI's decisions on next steps, progress tracking, and escalation increasingly reflective of their intent. Email-template carryover similarly improved alignment in execution, with participants reporting that drafts better matched their expectations over successive sessions. Quantitative measures supported this pattern: perceived value alignment via policies and templates increased from 5.50/7 to 6.12/7, drafting comfort rose from approximately 6.0 to nearly 7.0/7 across all email types, and effort dropped from 3.20/7 to 1.88/7 ($p = 0.054^{\wedge}$). P4 noted that edits to greetings, closings, and phrasing carried over into later drafts: ``\textit{Compared to the last [usage], DoubleAgents get my style\ldots{} so I edited fewer times this time.}'' P3 observed: ``\textit{I found that the agent's recent actions followed exactly the policies I had in mind --- super aligned.}''

\paragraph{\textbf{{\protect\tikz[baseline=-0.5ex] \protect\draw[fill=dot2, draw=black, line width=0.0mm] (0,0) circle (0.1cm);} Visualizations made the agent's reasoning legible, enabling confident evaluation.}}
The dashboard's temporal, social, and procedural views let participants verify what the agent was doing without parsing raw data. P6 found the communication visualization useful for showing full email exchange histories: ``\textit{Everything is consistent inside the system --- when some speakers were not followed up, I could map it back to the visualization and confirm accuracy\ldots{} [It] helped me build trust\ldots{} For tomorrow, I'll spend less time checking the email replies [inside the chat interface] one by one, and instead [just check] the Communication History on the side.}''

\paragraph{\textbf{{\protect\tikz[baseline=-0.5ex] \protect\draw[fill=dot1, draw=black, line width=0.0mm] (0,0) circle (0.1cm);} The coordination agent served as memory and reasoning partner, removing tracking burden.}}
By maintaining full state across days and weeks of coordination---who was contacted, who responded, what is pending---the agent freed participants from the cognitive overhead of tracking progress themselves. This let them focus on judgment and decision-making rather than bookkeeping. P7 noted that DoubleAgents was ``\textit{pretty good at figuring out plans, keeping track of who's responded, who's not, who's available, who's not, like what to do moving forward.}'' P1 similarly shared that ``\textit{now ...I didn't have to use my brain too much [to make sure things are consistent], just a little bit---so it handled the task really well.}''

\subsubsection{\textbf{\uline{Where did participants still need control rather than full delegation?}}}
\label{sec:control}

\medskip
\noindent 
While comfort with delegation grew, participants identified boundaries where human judgment remained essential.

\paragraph{\textbf{Stop hooks for flagging edge cases.}}
Participants highlighted stop-hook flagging as critical for maintaining safe delegation boundaries. P1 described issue flagging as ``\textit{one of the most important features of the system}'': ``\textit{[With the issue flagging,] I skim [through the other features] now\ldots{} It made me trust the system because I know it won't just execute something random\ldots{} It flags when it doesn't know\ldots{} That helps elicit my values and how I want to handle things --- and [then DoubleAgents] respects that\ldots{}}'' P8 described a case where flagging prevented a misinterpretation of progress: when they skimmed a reply and thought everything was resolved, the flag showed that it was not. Stop hooks gave participants confidence to offload routine execution precisely because they knew the system would pause at points of uncertainty. Comfort in offloading, in other words, was built not by the agent performing flawlessly, but by making its boundaries of confident action transparent and actionable.

\paragraph{\textbf{Clarification emails as a natural boundary for delegation.}}
The notable exception to increasing delegation comfort was clarification emails sent without user confirmation, which showed little change across sessions ($p>0.1$). P10 explained: ``\textit{For clarification emails, I might want to double-check\ldots{} [they're] usually not in the plan [or policies]\ldots{} I'm not comfortable [offloading that] to AI\ldots{}}'' Unlike routine emails that follow established templates, clarification emails require the agent to identify ambiguity, judge what information is missing, and formulate the right question --- all of which demand user-specific judgment that resists encoding into policies. This suggests a natural boundary for delegation: as agents take on more execution, the critical human role shifts toward decision-making at points of uncertainty, precisely the moments where knowing ``when and how to ask the user'' becomes the key alignment challenge.

\subsubsection{\textbf{\underline{Simulation as an accelerator for alignment}}}
\label{sec:simulation}
The three alignment mechanisms described above---policies, visualizations, and the coordination agent---improve through repeated interaction. But in live deployment, edge cases surface slowly: a speaker may take days to reply, and unusual requests arise only once across an entire series. Simulation accelerates this by compressing weeks of coordination into minutes while maintaining realistic diversity of cases. Participants found simulations realistic (5.90/7 to 6.75/7), useful for surfacing potential problems (5.80/7 to 6.50/7), and confidence in DoubleAgents robustness increased through simulation (5.80/7 to 6.38/7). P3 described:
``\textit{The simulations felt realistic and true to the person [they] were simulating\ldots{} it elicited possible problems that would happen in real life\ldots{} Like people asking for funding or needing to reschedule---it forced me to think about how I would handle that and to create relevant policies. I can trust the system to handle that in the future since it already happens in the simulation.}''

Simulation served all three mechanisms as a testbed: it provided a safe environment to probe whether agent behavior matched user intent (\textit{policies}), populated the dashboard with realistic state for practicing oversight (\textit{visualizations}), and surfaced ambiguous situations where the agent should pause (\textit{stop hooks})---letting participants define delegation boundaries in advance rather than discovering them through live mistakes. This parallels unit testing in software engineering, but where correctness is user-specific rather than universally defined. Simulation is not required---alignment can develop through live use---but it compresses the iteration cycle and lets participants build confidence before real social stakes are involved.
\section{Deployment Studies}
\label{sec:deploy}

\subsection{Study Design}
We conducted three real-world event-planning deployments (D1--D3) with two organizers (Q1, Q2), both PhD students (one male, one female; average age 31) with prior experience organizing academic events and using generative AI for email writing. 
Table~\ref{tab:deployments} summarizes the deployments.

\begin{table}[h]
\vspace{-2px}
\centering
\footnotesize
\begin{tabular}{cccc}
\toprule
\textbf{Deployment ID }& \textbf{D1} & \textbf{D2} & \textbf{D3} \\
\midrule
\textbf{Organizer ID} & Q1 & Q2 & Q1 \\
\textbf{Task} & Fall seminar & Workshop panel & Spring seminar \\
\textbf{Date} & Aug 2025 & Sep 2025 & Jan--Feb 2026 \\
\textbf{Candidate pool} & 23 & 18 & 20 \\
\textbf{Speakers invited} & 10 & 6 & 11 \\
\textbf{Deployment length} & 7 days & 4 days & 12 days \\
\bottomrule
\end{tabular}
\vspace{3px}
\caption{Summary of three real-world deployments.}
\label{tab:deployments}
\vspace{-20px}
\end{table}

After providing consent, organizers were introduced to DoubleAgents through a simulation-based demonstration of planning--execution loops. Before deployment, each organizer completed at least two simulation sessions (2+ hours) using personas based on real candidates. They refined policies, templates, and contextual inputs until the agent's proposed actions aligned with their expectations.

For live deployment, we disabled the respondent module and connected DoubleAgents to Gmail's SMTP service, enabling organizers to manage real email-based coordination tasks. After confirming sufficient alignment, organizers used the system throughout the deployment periods shown in Table~\ref{tab:deployments}. The agent operated on live emails while organizers retained approval authority over all actions. We logged system interactions, including email sends, scheduling updates, and policy adaptations, and collected qualitative feedback through regular check-ins.

\subsection{Deployment Results}
Both organizers reported that DoubleAgents fit seamlessly into email-heavy coordination workflows that conventional tools cannot manage, attributing their willingness to offload tasks to explicit, editable policies and templates that made the agent's reasoning legible. Q2 valued editable policies over AI tools that hide reasoning and remarked, "\textit{I trust AI [systems] for specific use cases more}." Q1 stressed that messages must make speakers "feel they are special" and found DoubleAgents generated strong starter drafts that improved over time. Interactive simulation helped establish alignment: both organizers used real invitee names and persona-specific edge cases to refine policies and templates. While alignment transferred to predictable scenarios, deployment revealed unanticipated dynamics requiring ongoing policy and tone adjustments; Q1 found renewed simulation necessary because ``\textit{real life is super complex.}''

DoubleAgents redistributed cognitive labor through three mechanisms: \textbf{\textit{{\protect\tikz[baseline=-0.5ex] \protect\draw[fill=dot2, draw=black, line width=0.0mm] (0,0) circle (0.1cm);} Visualization as shared group representation.}} Q1 found the color-coded visualization of speaker states useful enough to adopt a similar scheme for other spreadsheet-based tracking tasks, while Q2 said it supported trust through "\textit{natural mapping}" to his workflow. \textbf{\textit{{\protect\tikz[baseline=-0.5ex] \protect\draw[fill=dot3, draw=black, line width=0.0mm] (0,0) circle (0.1cm);} Policy and artifact refinement as externalized strategy.}} Both organizers iteratively edited policies and templates to improve alignment; Q1 adjusted follow-up frequency, formality, and tone, while Q2 valued explicit, editable rationales. These artifacts accumulated organizational knowledge over time, reducing review. \textbf{\textit{{\protect\tikz[baseline=-0.5ex] \protect\draw[fill=dot1, draw=black, line width=0.0mm] (0,0) circle (0.1cm);} Coordination agent as attention manager.}} DoubleAgents managed overlapping reply-and-assign cycles, letting Q2 check coordination status without scanning his inbox. Q1 said "\textit{DoubleAgents was really helpful and saved me a lot of time\ldots{} Policy-abiding [actions and template-following] emails are really useful}," but emphasized that ``\textit{The simulation part cannot be skipped.}''

\section{Limitations and Future Work}
\paragraph{Sample and study design.} Our lab study involved 10 participants in simulated scenarios, limiting generalizability, though the within-subjects design across two sessions yielded rich insights into how alignment strategies and delegation boundaries evolved with repeated use, and three week-long deployments validated these patterns in real-world settings. All participants were frequent generative-AI users; larger, more diverse samples, including users with less AI familiarity, would help clarify onboarding challenges and how simulation supports alignment across experience levels.

\paragraph{Task scope.} All of our evaluation is scoped to a single family of coordination tasks: seminar and event scheduling. Every policy, persona, seed rule, deployment, and technical evaluation instance is scheduling-specific, and the unpredictability our simulations surface (e.g., late replies, modality requests, date conflicts) is likewise scheduling-specific. Other socially embedded coordination settings we discuss as motivation in Appendix A, such as shift assignment and interview scheduling, share the same structural features — evolving state, social stakes, and negotiation over availability — but are not evaluated here. Whether the framework's three components, and the alignment artifacts users accumulate, transfer beyond scheduling-type coordination remains an open empirical question.

\paragraph{Simulation fidelity.} While our simulations reflected realistic coordination challenges, they cannot fully capture real-world unpredictability, and we validated them only through participant-rated realism rather than against real correspondence. Because lab-study personas were hand-authored by the research team, policies refined in simulation may partly reflect our assumptions about how invitees behave; deployment organizers mitigated this by building personas from real candidates. Grounding both the coordination agent and the respondent simulator in users' historical data — past emails, decisions, and communication patterns — could enable more personalized modeling of values and tone, and validating simulated replies against real ones is a natural next step.

\paragraph{Model dependence.} DoubleAgents assigns different models to different subtasks (Appendix~\ref{app:models}): a reasoning-optimized model for coordination planning, where deliberation quality matters most, and a faster general-purpose model for drafting, simulation, and edge-case detection, where per-call latency and cost dominate because these run many times per simulated day. Our technical evaluation varied state representation and shot count but not prompt phrasing or length, so sensitivity to prompt formatting is untested. Future models with stronger social reasoning and long-horizon planning may improve coordination quality, and tighter integration between user-specific data and model reasoning remains an open challenge.

\section{Conclusion}
We present DoubleAgents, an interactive alignment tool for human coordination through user intervention, value-reflecting policies, rich state visualizations, and uncertainty flagging. 
Technical evaluations validate key system capabilities of DoubleAgents, including policy selection and edge-case detection. A two-session user study (N=10) shows users' increased willingness to delegate both planning and execution, supported by policy-based alignment, interpretable visualizations, and explicit delegation boundaries. Three real-world deployments show that alignment learned through simulation transfers to live coordination while continuing to evolve in practice. Our findings frame alignment not as a one-time specification problem but as an iterative process, and position distributed cognition as a principled approach for enabling progressive delegation with retained user control.

\begin{acks}

    This work was supported by the Columbia Center for AI and Responsible Financial Innovation (CAIRFI). We would like to thank people from the Data, Agents, and Processes Lab (DAPLab) at Columbia University for their valuable discussions on the paper. We also thank all reviewers for their constructive feedback and
suggestions, which significantly improve our work.
\end{acks}

\bibliographystyle{ACM-Reference-Format}
\bibliography{bib}


\appendix

\section{Seminar Planning: A Human Coordination Problem}
\label{sec:background}

Human coordination—such as assigning work shifts or scheduling job interviews—is both cognitively demanding and socially delicate. Unlike tasks with stable inputs and predictable outcomes, coordination requires managing uncertainty, reconciling multiple perspectives, and sustaining relationships over time. Even small lapses—missed follow-ups, unclear tone, or misaligned expectations—derail coordination and strain relationships. These challenges make coordination both essential and persistently difficult in many areas of knowledge work.

We study these challenges through one concrete and consequential setting: organizing an academic seminar series. A typical series spans an entire semester and involves inviting a new external speaker nearly every week, amounting to 8–12 speakers per term. Early on, organizers must identify potential speakers, gather their availability constraints, and balance their own preferences-such as ensuring topical diversity across weeks and securing highly sought-after senior researchers who are harder to schedule.

However, the most unpredictable and challenging part is dealing with responses. Some people do not reply at all and require multiple follow-ups to elicit a response. Some give only vague and tentative replies that need repeated confirmation. Some can be flaky and unreliable, offering slots but later canceling. And many unexpected problem cases arise: people might be on parental leave, ask for visa letters or funding support, recommend a different speaker, request a different date or talk modality, or raise logistical demands. In such human coordination tasks, you can rarely predict what is coming. Therefore, conventional scheduling tools such as Calendly\footnote{https://calendly.com/} or When2Meet\footnote{https://www.when2meet.com/} are ill-suited to this setting. They assume stable constraints and one-shot coordination, yet seminar scheduling requires ongoing negotiation, persuasion, and relationship maintenance. Unlike ordinary meeting scheduling, seminar speakers are doing you a favor, which means you cannot simply send them a link to pick a time—you must persuade and negotiate. Social dynamics and etiquette also play a large role: organizers must time their emails carefully to avoid overwhelming, annoying, or losing momentum with potential speakers, who are often busy scholars. Knowing when to nudge, follow up, or wait is subtle. Email remains the standard medium for professional and academic outreach, allowing organizers to adapt tone—formal, casual, or enthusiastic—to different speakers in ways scheduling links cannot. It also supports sharing rich context, such as talk topics, background, past speakers, or travel plans—far beyond simply choosing a calendar slot.

Thus, experienced organizers adopt personal strategies to cope with such a demanding task. For example, some follow a rule of “wait two days, then follow up” to nudge busy people who may not read emails right away, while others wait a full week before reaching out again. Some prefer contacting close collaborators or junior researchers first, while others prioritize securing prominent professors or distant speakers early to reduce travel constraints. When conflicts arise between speakers, organizers might “give faculty priority and ask students to be more flexible,” carefully deciding whom to approach for alternative availability to minimize disruption. For non-responders or last-minute cancellations, they must judge when and how often to follow up before moving on to new options—and sometimes even decide when it is appropriate to “beg” busy invitees for additional availability. Because plans are constantly evolving, organizers become locked in a cognitively and time-intensive loop of tracking shifting states, replanning, and executing revised plans as new information arrives—yet they often cannot proceed with the next steps or finalize allocations until replies are received. Since seminar scheduling requires nuanced, context-dependent decisions that reflect each organizer’s values and preferences, it is difficult to fully automate. An effective seminar scheduling tool should not attempt to handle the process independently; instead, it must learn from and adapt to the organizer, respecting social norms, relational dynamics, and the inherent unpredictability of responses. 

\section{System Implementation Details}
\label{sec:appendix_implementation}


We present DoubleAgents, a system for aligning agentic AI through interactive simulation. Built on the ReAct framework~\cite{yao2023react}, DoubleAgents supports iterative planning and execution via a team of AI agents that collaborate with users on complex coordination tasks. DoubleAgents takes as input a user-defined goal (e.g., ``Organize a seminar with four speakers and four slots''), persona descriptions for simulated respondent agents, the user's existing email templates, and a set of generic scheduling policies. Through iterative simulations, the system produces a finalized assignment of speakers to slots, along with refined policies, updated email templates, and executable stop hooks that users can later deploy in real-world agentic workflows.

The system architecture comprises four key components: (1) a \textit{Simulation Engine} that drives the iterative planning-execution loop through coordinated AI agents (Section~\ref{subsec:simulation}), (2) a \textit{Policy Module} that governs agent reasoning and adapts to user preferences (Section~\ref{subsec:policy}), (3) an \textit{Edge Case Detector} that identifies situations beyond existing policy coverage and escalates them for user intervention (Section~\ref{subsec:edge_case}), and (4) a \textit{Dashboard Visualization} that provides real-time visual feedback on task state and progress (Section~\ref{subsec:dashboard}).

These four components interact in a loop. The \textit{Policy Module} supplies policies to the coordination agent inside the \textit{Simulation Engine}, which in turn exchanges emails with simulated respondent agents to advance the scheduling task. When a respondent's reply falls outside existing policy coverage, the \textit{Edge Case Detector} escalates the issue to the user and feeds the resulting clarification back into the Policy Module as an updated policy. Throughout this cycle, the \textit{Dashboard Visualization} renders the evolving task state so that the user can monitor progress and intervene at any point. We describe each component in detail below.

\subsection{Simulation Engine}
\label{subsec:simulation}

The Simulation Engine is the core of DoubleAgents, implementing the iterative planning-execution loop that drives the coordination process. It comprises three main subcomponents: the \textit{Coordination Agent}, which reasons over the task and orchestrates actions; the \textit{Simulated Respondent Agents}, which emulate human behavior to create a realistic environment; and the \textit{Context Management Module}, which maintains the evolving problem state.

\subsubsection{Coordination Agent}
\begin{figure*}[]
\centering
\begin{subfigure}{0.3\textwidth}
  \centering
  \includegraphics[width=\linewidth]{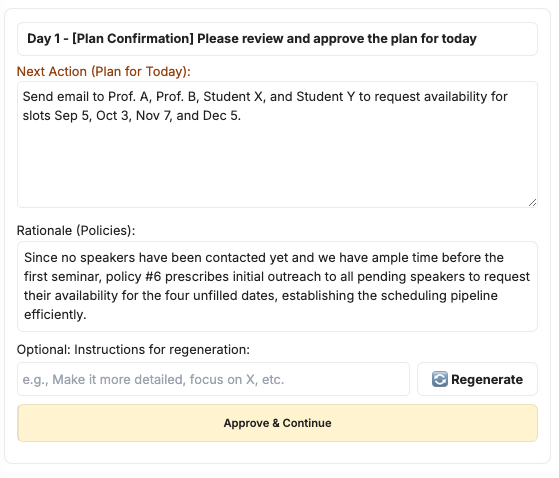}
  \caption{The pop-up window for a generated plan.}
  \label{fig:plan_gen}
\end{subfigure}%
\hfill
\begin{subfigure}{0.3\textwidth}
  \centering
  \includegraphics[width=\linewidth]{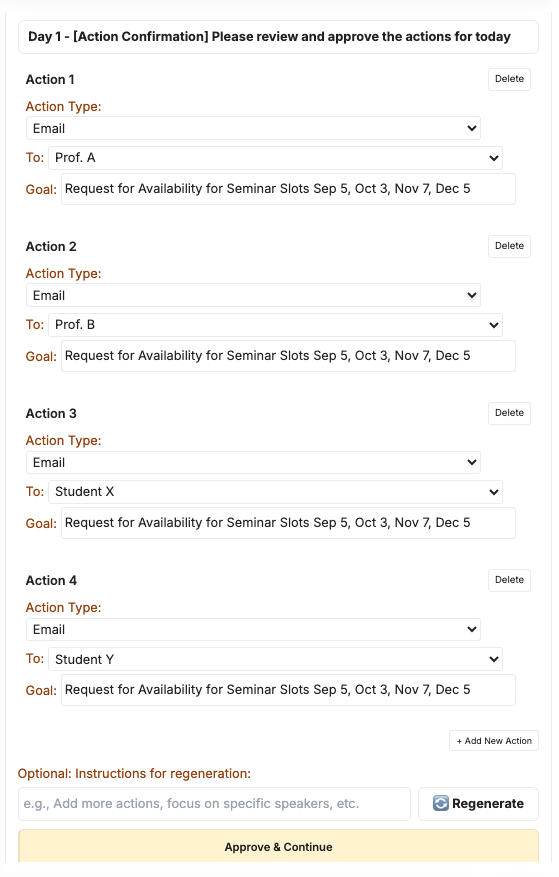}
  \caption{The pop-up window for generated actions following the plan.}
  \label{fig:action_gen}
\end{subfigure}
\hfill
\begin{subfigure}{0.3\textwidth}
  \centering
  \includegraphics[width=\linewidth]{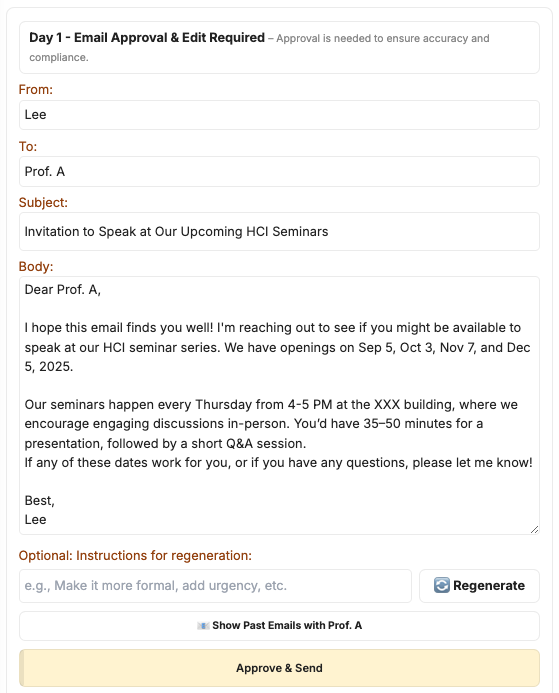}
  \caption{The pop-up window for a generated email.}
  \label{fig:email_gen}
\end{subfigure}
\caption{Examples for plan and action generation.}
\label{fig:plan_action_email_gen}
\end{figure*}

The coordination agent is the primary reasoning component of the system, responsible for assessing the current problem state, selecting relevant policies, generating plans, and executing actions. It operates within the ReAct framework and proceeds through the following stages at each simulation step:
\begin{itemize}
    \item \textbf{Progress Understanding.} The agent reviews the current problem context, including speaker information, email logs from previous days, and the history of user-approved plans and actions, and generates a concise summary of the current status. This summary indicates which days have been scheduled, what actions have been taken, and the most recent contact date for each speaker. The prompt used for generating the progress summary is provided in Appendix~\ref{subsubsec:coordination-progress-summary}.
    \item \textbf{Policy Selection.} Based on the progress summary, the agent invokes the policy selection tool to identify the most relevant policies for the current state (Section~\ref{subsec:policy}).
    \item \textbf{Plan Generation.} Guided by the selected policies, the agent proposes a high-level plan (e.g., ``Send email to all speakers to request availability for all seminar slots''). The plan and its rationale are presented to the user via the Interactive Interface (Figure~\ref{fig:screen_all}(Middle)) for review and approval. Users may approve, edit, or request regeneration of the plan (Figure~\ref{fig:plan_gen}).
    \item \textbf{Action Generation.} Upon plan approval, the agent generates a set of concrete actions to operationalize the plan (Figure~\ref{fig:action_gen}). There are three types of actions: (a) drafting emails to selected respondents, (b) sending emails to specified recipients, and (c) awaiting responses from certain respondents. Users retain full control over actions and may edit, remove, or request regeneration. Detailed prompts for plan and action generation are provided in Appendix~\ref{subsubsec:coordination-action-generation}.
    \item \textbf{Action Execution.} The agent instantiates each approved action by invoking the appropriate tools from its toolbox (Figure~\ref{fig:system}). For email drafting, it calls the \texttt{writeEmail()} API, which generates emails given the current problem state, seminar details, user-provided templates, and historical email edits. Email drafting is performed using GPT-4o. For sending, it calls the \texttt{sendEmail()} API targeting the appropriate speaker email addresses. For delayed responses, it calls the \texttt{waitResponse()} tool, pausing execution and advancing to the next pending action or simulated day.
\end{itemize}

Users interact with the coordination agent through the Interactive Interface (Figure~\ref{fig:screen_all}(Middle)), which displays the full history of plans, actions, and emails as interactive pop-up windows. The interface also supports natural language queries (e.g., ``\textit{What did Prof. Lucy Tran reply in the last email?}''), powered by GPT-4o (see Appendix~\ref{subsec:freeform-qa-chat}). We use GPT-o3 as the backbone of the coordination agent, given its strong reasoning capabilities.

\subsubsection{Simulated Respondent Agents}
Drawing inspiration from \cite{park2023generative}, we employ LLMs to simulate each respondent agent, generating actions that align with their assigned personas. Upon receiving emails from the coordination agent, each simulated respondent first reasons over its persona and then generates an appropriate action. For instance, a respondent whose persona specifies ``\textit{You are very busy and reply very late. Normally, it takes you three follow-up emails to respond}'' will delay its reply accordingly, while a respondent described as prompt in communication will reply immediately. If a respondent's persona includes specific preferences (e.g., ``prefers presenting over Zoom''), the generated response will reflect those preferences. Each respondent composes and sends email responses, which advance the scheduling process and enable the user to coordinate subsequent steps. The prompt for the speaker agent is detailed in Appendix~\ref{subsec:simulated-respondents}. We adopt GPT-4o for email response generation via the \texttt{sendEmail()} tool.

\subsubsection{Context Management Module}
\begin{figure*}[!h]
\centering
        \includegraphics[width=.8\linewidth]{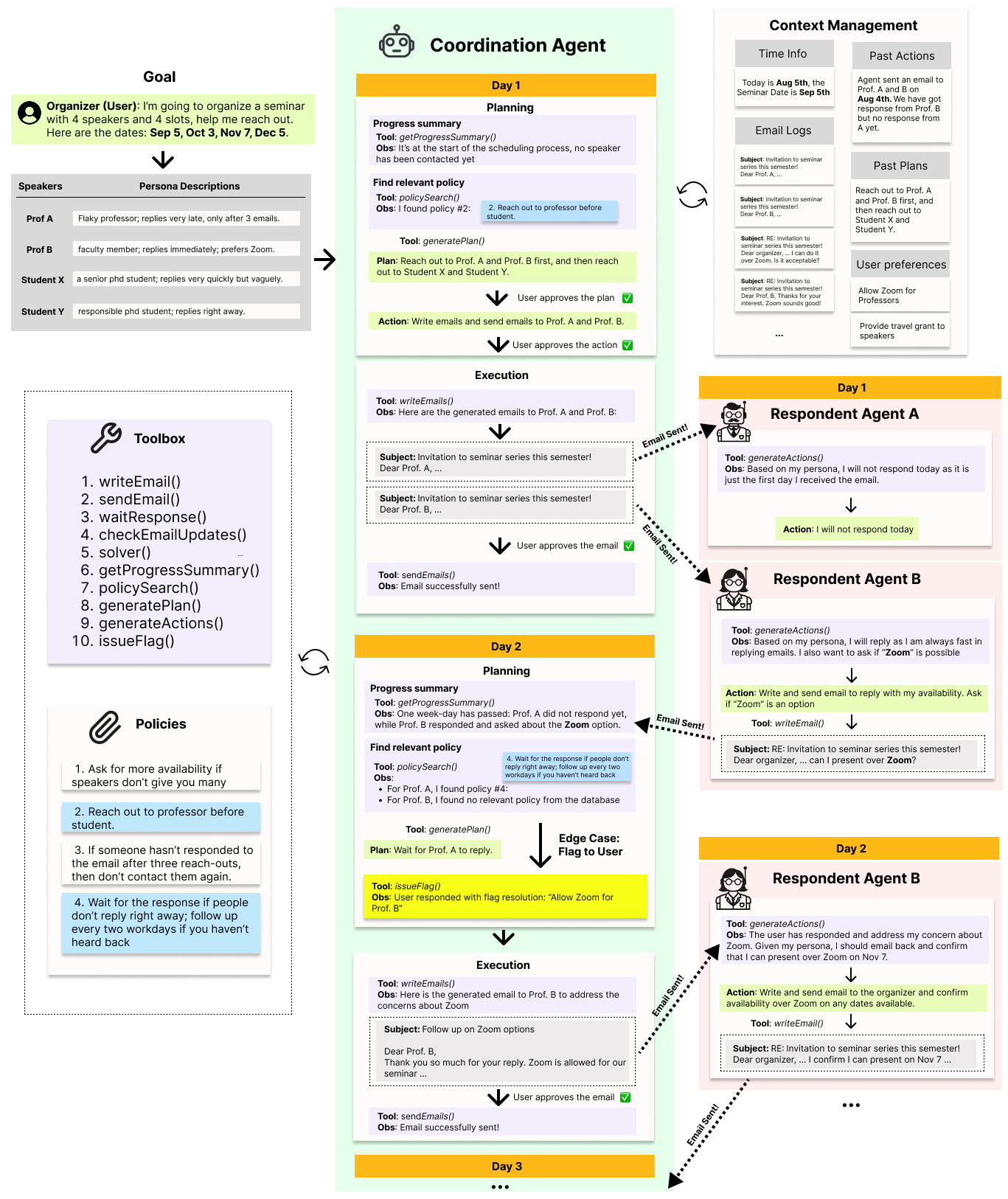}
    \caption{System diagram of DoubleAgents illustrating a day-by-day ReAct workflow that couples policy-guided planning with human oversight and LLM simulation. \textbf{Inputs}  (top left) include the organizer’s goal, seminar slots, speaker personas, and a toolbox of callable functions. The \textbf{coordination agent} (center) iteratively: summarizes state; selects applicable policies; proposes a plan and action that the user can Regenerate or Approve; executes by drafting and sending emails, or waiting for responses. If a reply falls outside policy coverage, issueFlag escalates an edge case to the user for clarification before proceeding. The \textbf{context management} (right) maintains time information, past plans/actions, email logs, user preferences, and persona data, continuously updating to ground subsequent steps. \textbf{Simulated respondent agents} (bottom right) generate realistic, persona-consistent behaviors and replies that drive the loop across days. For a detailed walkthrough, please refer to Section \ref{subsec:walkthrough}.
}
  \Description{Diagram showing the architecture of the DoubleAgents system, with components for goal setting, agent coordination, context management (including edge case detection), and various submodules such as toolbox, orchestrator, and simulated respondents.}
  \label{fig:architecture}
    \label{fig:system}
\end{figure*}

The Context Management Module maintains all relevant contextual information throughout the scheduling process. DoubleAgents represents the problem context using JSON structures, where each key denotes a specific type of contextual information (e.g., speaker profiles, time constraints, email history) and each value contains the corresponding content. As shown in Figure~\ref{fig:system}, the problem context includes keys such as ``time info'' (reflecting seminar slots), speaker details, and communication logs. The context is dynamically updated at each step, incorporating user feedback, agent actions, and simulated responses to maintain an up-to-date representation of the task state.

\subsection{Policy Module}
\label{subsec:policy}
The Policy Module serves as the central mechanism through which user values and preferences are encoded and operationalized within the system. Policies are natural language rules that guide the coordination agent's reasoning and planning at each step of the scheduling process. They are displayed in the Policy Panel (Figure~\ref{fig:screen_all}(Left)), where users can inspect, add, edit, or delete them at any time.

\subsubsection{Seed Policies}
To bootstrap the system with broadly applicable guidelines, the lead authors designed a set of six initial “seed policies'' that address common scheduling challenges: (1) individuals being reluctant to share their full availability, (2) ambiguous or unclear responses, (3) the need for multiple follow-ups to elicit responses, (4) resolving scheduling conflicts between two speakers, (5) initiating the process by broadly reaching out to potential participants, and (6) determining when to cease follow-up attempts with non-responsive participants. Each policy is designed to address a specific challenge and serves as a default guideline that the coordination agent can draw upon when generating plans.

\subsubsection{Policy Selection}
At each planning step, the coordination agent invokes a policy selection tool that identifies the most relevant policies given the current problem state. For instance, at the beginning of the scheduling process when no speaker availabilities have been collected, the system selects the policy: ``\textit{If we are just beginning and have not yet contacted any speakers for availability, then ask each speaker for their availability across all slots.}'' The selected policy is highlighted in the Policy Panel (Figure~\ref{fig:screen_all}(A)), providing users with a transparent rationale for the agent's subsequent plan. Details of the prompts used for policy selection are provided in the Appendix~\ref{subsubsec:coordination-policy-selection}.

\subsubsection{Policy Refinement via Stop Hooks}
Policies are not static; they evolve throughout the simulation process. When the Edge Case Detector (Section~\ref{subsec:edge_case}) identifies a situation not covered by existing policies and the user provides a clarification, the system records this intervention as a \textit{stop hook} and uses it to refine the policy set. For example, if a speaker requests to present via Zoom and the user specifies a preference for in-person attendance, this clarification is converted into a new policy explicitly stating that remote participation is not preferred. In this way, user interventions during edge cases are transformed into reusable system policies that persist across subsequent simulation steps and can be deployed in real-world workflows.

\subsubsection{Implicit User Preference Learning}
Beyond explicit policy refinement, the system also supports implicit learning of user preferences by observing user-initiated edits to system-generated plans, actions, or emails. The coordination agent summarizes each edit into a concise statement, logs it in a user edit history, and incorporates it into the problem context for subsequent planning steps. This enables the agent to evolve its behavior in alignment with user preferences over time without requiring explicit policy authoring. We leverage GPT-4o for the summarization process, with prompt details provided in Appendix~\ref{app:imp_learn}.

\subsection{Edge Case Detector}
\label{subsec:edge_case}
The Edge Case Detector serves as a safeguard that identifies situations falling outside the coverage of existing policies and escalates them to the user for clarification. It is implemented as a tool that the coordination agent invokes to check whether an incoming message from a respondent agent can be addressed using the current policy set.

When the detector identifies an edge case---for example, a speaker requesting to present via Zoom when no existing policy addresses remote participation---it triggers a clarification request through the Interactive Chat Interface. The system presents the user with a concise description of the situation and a question prompting guidance (e.g., ``\textit{Prof. Lucy Tran requests a Zoom option due to circumstances preventing in-person attendance. Should I confirm Zoom availability for their seminar or insist on in-person participation?}''). The user then provides a natural language instruction (e.g., ``\textit{Tell Prof. Lucy Tran we prefer in-person due to the potentially large audience}''), which the coordination agent processes to generate an appropriate action.

Critically, each user clarification is recorded as a \textit{stop hook} and used to refine the policy set (Section~\ref{subsec:policy}). This mechanism ensures that (1) ambiguous or novel situations are never resolved autonomously without user input, (2) user interventions are preserved as reusable policies for future reference, and (3) the policy set grows iteratively to cover an expanding range of scenarios. The edge case detector is implemented with GPT-4o, with prompts detailed in Appendix~\ref{subsec:edge-case-detection}.

\subsection{Dashboard Visualization}
\label{subsec:dashboard}

To support user understanding and foster alignment, DoubleAgents provides real-time visual feedback on the task state and system progress throughout the scheduling process. We implement a Flask-based dashboard interface designed to make the planning workflow interactive, legible, and actionable. The dashboard is composed of several layered visual components, shown in Figure~\ref{fig:screen_all} (Right):
\begin{itemize}
    \item \textbf{Assignment Progress Tracker.} This panel visualizes tentative and confirmed speaker-to-slot assignments. It highlights which seminar slots remain unfilled and which speakers have been successfully scheduled, enabling users to assess how close the current plan is to completion at a glance.
    \item \textbf{Calendar View.} Anchored in simulated days, this temporal layer helps users reason about timing, deadlines, and follow-up actions. The calendar highlights both the current simulated day and the seminar dates, allowing prioritization of actions based on urgency and scheduling constraints.
    \item \textbf{Communication History.} This panel displays the message history between the coordination agent and each speaker, offering a transparent view of the outreach and response status. It enables users to track ongoing communications, detect gaps, and assess follow-up needs.
\end{itemize}

These visual layers operate in parallel, allowing the system to efficiently track and update the state of each element in real time. At the end of each simulated day, the system generates a daily summary that highlights key events and decisions, using GPT-4o (with prompt details in Appendix~\ref{subsec:daily-summary}). The summary is presented within the Interactive Interface, providing users with a consolidated overview before advancing to the next simulated day.


\subsection{Model Configuration}
\label{app:models}

Table \ref{tab:model_config} summarizes the model used for each component. We select models by task demand: coordination planning requires multi-step reasoning over accumulated state and uses \texttt{GPT-o3}, whereas email drafting, respondent simulation, and edge-case detection are shorter-horizon generation and classification tasks that run several times per simulated day, making a faster and cheaper model preferable.

\begin{table}[t]
\centering
\small
\begin{tabular}{@{}llc@{}}
\toprule
\textbf{Component} & \textbf{Model} & \textbf{Sec.} \\
\midrule
\multicolumn{3}{@{}l}{\textit{Deployed system}} \\
Coordination agent (planning, actions) & o3 & B.1.1 \\
Email drafting (\texttt{writeEmail()}) & GPT-4o & B.1.1 \\
Simulated respondents (\texttt{sendEmail()}) & GPT-4o & B.1.2 \\
Free-form QA chat & GPT-4o & E.4 \\
Implicit preference learning & GPT-4o & B.2.4 \\
Edge case detector & GPT-4o & B.3 \\
Daily summary & GPT-4o & B.4 \\
\midrule
\multicolumn{3}{@{}l}{\textit{Technical evaluation only}} \\
Policy selection & o4-mini & C.1 \\
Edge case detection & GPT-4o & C.2 \\
Edge case test-suite generation & GPT-5 & C.2 \\
\bottomrule
\end{tabular}
\caption{Model assignment by component. Models in the lower block are used only for the technical evaluation (Appendix~\ref{app:tech-eval}) and are not part of the deployed system.}
\label{tab:model_config}
\end{table}
\section{Technical Evaluation Setup}
\label{app:tech-eval}

This appendix details the experimental setup for the two technical evaluations reported in Section~\ref{tech-eval}.

\subsection{Policy Selection}
Policy selection refers to identifying coordination policies that guide the generation of high-quality, user-aligned plans. Given a particular problem state (e.g., unfilled time slots and unresponsive speakers), the policy selection module retrieves relevant policies that can inform the next planning step. For example, if no speakers have been contacted yet, a suitable policy would be: “\textit{If we are just beginning and have not yet contacted any speakers for availability, then ask each speaker for their availability across all slots.}” In our implementation, we adopt a seed set of general-purpose coordination policies curated by domain experts.

To study how different representations of the problem state affect policy selection, we compare four input conditions that vary in the level of abstraction and context provided to the model:
\begin{itemize}[left=0pt, labelsep=5pt, itemsep=2pt]
    \item \textbf{Speaker-only} context, which includes  information about each speaker’s persona. For example: “\textit{Speaker 1 (Prof. Adrian Smith):  busy, often replies only after multiple follow-ups; Speaker 2 (Dr. Lucy Tran): usually responds within 1–2 days; Speaker 3 (Sarah Malik): highly responsive and flexible; Speaker 4 (Michael Chen): prefers early scheduling but has not replied recently.}”.
    \item \textbf{TimeSlot-only} context, which includes the current status of all time slots and their constraints. For example: “\textit{Four slots: Sep 5 (unfilled), Oct 3 (filled), Nov 7 (filled), Dec 5 (unfilled). The Nov 7 slot has been confirmed with a speaker, while Oct 3 is tentatively held. Remaining open slots require confirmation.}”
    \item Combined \textbf{TimeSlot and Speaker} context, which includes both sources of information without additional abstraction. For example: “\textit{Speaker 1 (busy, slow to reply) and Speaker 2 (moderately responsive) were last contacted two days ago; Speaker 3 (responsive) has not yet been contacted; Speaker 4 prefers early slots but has not replied. Slot status: Oct 3 and Nov 7 are filled (Nov 7 confirmed), while Sep 5 and Dec 5 remain open.}”
    \item \textbf{Summary-based} context, which includes the information of the speakers, time slots, and the summary of the current task state to provide a structured, high-level snapshot of the system. An example summary reads: “\textit{One slot has been secured: speaker 4 confirmed for Nov 7 (2025-11-07) and received a confirmation email on 2025-08-07. All other speakers are still pending. Last outreach dates: speaker 1 and speaker 2 were last emailed on 2025-08-08; speaker 3 was last emailed on 2025-08-07.}”
\end{itemize}

Prompt templates for all conditions are provided in Appendix~\ref{sec:technical-eval}. We construct a test suite of 30 examples sourced from real user study sessions, covering a diverse set of scheduling scenarios. For each example, we treat the set of policies approved by users during the study as ground truth. All evaluations are conducted using the {\small\texttt{o4-mini}} model with a default temperature of 1, and we report the average F1 score across three independent runs for each condition.

\subsection{Edge Case Detection}
Edge case detection aims to determine whether a respondent’s message introduces a scenario that falls outside the scope of existing coordination policies. For instance, if a speaker requests to present via Zoom while no policy addresses virtual presentations, the system should flag this case for user clarification and further instruction.

To study how different prompting strategies affect detection performance, we compare three conditions that vary in the level of guidance and policy awareness provided to the model: (1) \textit{zero-shot prompting}, where the model is given only the task instruction and the incoming message without any examples; (2) \textit{few-shot prompting without policy context} (\textbf{few-shot w/o policy}), where the model is provided with representative examples of edge and non-edge cases but does not have access to the current set of coordination policies; and (3) \textit{few-shot prompting with policy context} (\textbf{few-shot w/ policy}), where the model is given both demonstration examples and the existing policies, allowing it to assess whether a new situation is already covered or requires escalation.

Prompt templates for all conditions are provided in Appendix~\ref{sec:technical-eval2}. We construct a test suite of 100 examples, manually validated by the lead authors to ensure diversity and coverage across common coordination scenarios and edge cases. The construction methodology is detailed in Appendix~\ref{sec:technical-eval2}. All experiments are conducted using {\small\texttt{GPT-4o}}, with temperature set to 0.7 and a maximum token limit of 100. We report the average accuracy across three independent runs for each condition.

\section{User Study Procedures and Data Analysis}
\label{sec:userstudyProcedures}
\subsubsection*{Test Scenario Preparation.}

We selected eight distinct personas identified by three organizing experts, representing realistic and diverse roles, research areas, and backgrounds (e.g., a fifth-year PhD student studying VR currently on the job market; a busy junior professor in NLP who spends much of her time on her startup). For each persona, we specified reply behaviors (responsiveness, email writing style; e.g., slow responder, writes only short emails) and optional issues (potential concerns they might raise with speakers; e.g., requesting additional funding even though their in-person visit is already funded). We split them into two groups and created seminar organizing instances. We set the timeline as a monthly seminar schedule, with the first seminar one month after the simulation start date.

\subsubsection*{Day 1, Preparation and Setup (30min)} Participants first provided informed consent and completed an AI comfortableness survey assessing their perceptions and expectations with AI automation tools, human coordination tasks, and demographics. They also edited a few seeded templated seminars organizing emails according to their tone to feed the system their writing style. They were then introduced to the DoubleAgents interface and task. One researcher gave a brief walkthrough using a visual diagram and demonstration, covering the system’s core components, including the planning-execution coordination agent loop, respondent agent simulation, policy and context modules, visualization, and the flag mechanism. 

\subsubsection*{{Day 1, First System Usage and Evaluation (60min)}} Participants then completed a seminar scheduling task (40 minutes), interacting with four simulated speakers with diverse personas, response behaviors, constraints, and coordination tensions. During the task, participants articulated and shared their mental models and usage strategies for DoubleAgents. Afterward, they completed another time of AI comfortableness survey and a system experience survey assessing experience, workload, and attitudes toward specific components such as control, usefulness, and trust. Survey questions were adapted from NASA TLX~\cite{hart2006nasa} and prior AI perception, reliance, and long-term evaluation studies~\cite{Ross, 10.1145/3715275.3732045, taolongitudinal}. Finally, a short interview asked participants to reflect on their overall experience, including any insights or policies they would like the system to incorporate before Day 2.

\subsubsection*{Day 2, Second System Usage and Evaluation (60min)} Before users’ next DoubleAgent usages, the researcher first ensured that the system incorporated participants’ first-day planning context and any preferences they had established via policies or system edits. Participants then completed another seminar scheduling task (40 minutes), followed by the same AI perception survey, the system experience survey, and a short interview. Day 2 interviews included follow-up questions comparing participants’ mental models, perceptions, experiences, and expectations of the system between Day 1 and Day 2.

\subsubsection*{Data Collection and Analysis} We logged participants’ interactions with AI suggestions, including regenerations, rejections, and edits, which were incorporated into DoubleAgents’ implicit learning module and our analysis. Interview data were analyzed using thematic analysis to understand how users perceive the usefulness and trust drive about each feature, and quantitative data were analyzed using paired-samples t-tests to examine whether participants’ experiences and perceptions of the DoubleAgents and AI tools across usage and days.

\section{DoubleAgents Prompts}
\label{sec:doubleagents-prompts}

\subsection{Coordination Agent}
\label{subsec:coordination-agent}

\subsubsection{Progress Summary}
\label{subsubsec:coordination-progress-summary}

\hfill
\begin{promptbox}
You are an expert seminar organizer assistant.  
Your task is to analyze the current progress and return a JSON object.

Input information may include:
- Speaker details (persona and description)  
- Possible dates  
- Scheduling rules  
- Email log  
- History of approved plans/actions  
- Time context  
- Any user suggestions  

You should mainly look through Email log and history of approved plans/actions to analyze the current progress. Do not assume any additional actions have been taken. 

Your JSON output should contain:
- summary: A concise summary of the current state, including which days are scheduled, what actions have been taken, and the last date we reached out to each speaker (for tracking and follow-up).  
- un\_or\_filled\_slots: Two lists of dates — one for filled slots and one for unfilled slots.  
- pending\_speakers: A list of speaker names who have not yet responded or are pending (how many days since we last reached out to them).  

When mentioning speakers, use the format “English Name (ID)”.
Avoid vague terms like *all speakers* or *pending speakers*.

----------

[Task Rules]

[Possible Dates]

[Time Context]

[Speaker Information]

Here are the history of Approved Plans and Actions:

[Approved Plans and Actions]

--------------------------------

Here is the history of past user edits and implicit learning (recent, action edits only):
[Implicit Learning]

--------------------------------

Past Email Log:
[Email Log]

\end{promptbox}

\subsubsection{Policy Selection}
\label{subsubsec:coordination-policy-selection}

\hfill
\begin{promptbox}
You are a helpful assistant. Given the following main reasoning result and context, select which of the following coordination policies should be applied. The chosen policies should be 100\% applied to the scenario! If you found that we didn't do anything yet from the reasoning, you should choose the 'we are just beginning' policy

Return a JSON list of the indices of the applicable policies (e.g., [1,2]) or [1]. If none apply, return an empty list [].

[Main Reasoning Result]

[Coordination Policies]

Return ONLY the JSON list of indices.

[Implicit Learning/User Edit History (recent, action edits only)]

\end{promptbox}

\subsubsection{Action Generation}
\label{subsubsec:coordination-action-generation}

\hfill
\begin{promptbox}

        You are an expert seminar organizer assistant.
        
        Your task: Given the organizing progress summary, the selected coordination policies, and the current time context, generate a concise plan for what must be accomplished today only.

        Output format:
        Return only one single string that contains:
        1. The plan (concise, no bullet points, no unnecessary multiple sentences).
        2. A short rationale starting exactly with "Why this plan? ...". The rationale should be about 40 words explaining why this plan was chosen according to the task progress summary context and how the policies/time context informed the decision.

        Example:
        "Send email to X to request availability for slots XXX. Why this plan? As we are just beginning the scheduling process and have not reached out to X yet, we followed policy \#6 (highlighted on the left) to obtain X's availability for the slots. This will help jumpstart the process.""

        Guidelines:
        - Use each speaker’s real English name.
        - Focus strictly on today (ignore tomorrow or later).
        - Valid actions (choose strategically):
        1. "Send email to X to request availability for slots XXX" (initial outreach).
        2. "Follow up with X" (if email sent but no reply).
        3. "Wait for response from X" (if email sent and still within expected waiting window).
        4. "Confirm and assign slot for X" (only if X has shared availability; avoid finalizing too early unless time is urgent).
        5. "Send clarification email to X" (if clarification on dates/details is needed).
        - If multiple actions are needed, combine them into one clear sentence (e.g., "Send email to X, Y, and Z to request availability for slots A, B, and C.").
        - Do not include any other tasks (e.g., slides, addresses, future prep).
        - Remember what day today is and when the last emailing was done (relevant for some policies).
        - Consider time urgency:
        - If few days remain, prioritize immediate action.
        - If weekend, default to waiting for responses unless a policy specifies otherwise.

        --------------------------

 Time Context:
 
[Time Context]

 Task Progress Summary:
 
[Progress Summary]

 --------------------------

 Applied Coordination Policies:
 
 [Applied Policy]

 --------------------------

 Implicit Learning/User Edit History (recent, action edits only):
 
[Implicit Learning - Action Edits]

--------------------------

\end{promptbox}

\subsubsection{Email Drafting}
\label{subsubsec:coordination-email-drafting}

\hfill
\begin{promptbox}
I am [Organizer Name], a seminar/workshop organizer. 
        Please help me write an email to [Speaker Name] about [Email Goal]. Please make sure it conveys that.
        Try to use a casual tone. If I haven't given you any details, keep it concise 
        and avoid unnecessary formalities.
        Sometimes it is a full email to invite some, while sometimes it might be just a followup email or a clarification / confirmation email. 
        So you should write different emails according to the situation and past email history.

        Please return a JSON object with the enhanced email containing:
        - "from": organizer (just organizer, not names),
        - "to": recipient (ensure it's just speaker's Full name, no other information, dont change)
        - "subject": subject
        - "body": enhanced email body that is professional and clear

        Make sure the email is natural language, professional, and incorporates all context provided.
        Consider the time urgency when writing the email. If only a few days remain until the seminar dates, convey urgency appropriately. 
        For follow-up emails (like following up / reminder emails -- no need to mention the seminar information.)
        You need to explicitly mention the dates, don't say "next seminar/next Thursday" or "Nov 1, 8, 10". You need to list the dates individually.
        Start the greeting with "Dear [Speaker name]".
        ---------------------------
        
        Here are some information of all the speakers we want to invite
        Context: [Progress Summary]

 Past Email History with that Speaker:
        [Past Email History with Speaker Name]
        ---------------------------

 Time Context:
    [Time Context]
        ---------------------------

 Seminar Information (can be opted out if this is not the first email)
       [Seminar Information]
        ---------------------------

 Users' template examples for invitation, if user has provided some, please try to follow them:
        
User's history invitation email :
    [Invitation Email Template]
    
User's history followup email:
   [Invitation Followup Template]
     
User's history confirmation emails :
 [Invitation Confirmation Template]
    
        ---------------------------

 When necessary, you should adapt some style edit that I have already taken on previous emails. However, be smart on whether they are the correct people to do so. 
        
[Implicit Learning - Email Edits]
\end{promptbox}

\subsubsection{Implicit User Preference Learning}
\label{app:imp_learn}
\paragraph{Plan and Action Edits}
\hfill
\begin{promptbox}
You need to describe the user’s edits by comparing the AI/system \{item\_type\} with the user-edited \{item\_type\}, making use of the history context.
        Output only a brief and concise summary of the edits. Do not infer motives. 
        Include relevant background context from the history — for example, when the change happened (if indicated) and how the text was modified. 
        Format: 'The user deleted XXX or added XXX to the \{item\_type\} when [context about the stage or progress of the allocation].' 
        Example: The user removed Tracy and Jack from the AI-generated outreach plan but kept others, such as Tanya and Adam, who are both faculty members.

Inputs:
history\_context: \{history\_context\}"
"ai\_text: \{ai\_text\}"
"user\_text: \{user\_text\}"
\end{promptbox}

\paragraph{Email Edits}
\hfill
\begin{promptbox}
You need to describe the user’s email edits related to format and style by comparing the AI/system {item\_type} with the user-edited {item\_type}, making use of the history context.
        You need to first analyze if the user edits relate to format and style. If not, you should return "No learning needed as the user edits are not related to format and style." Note that you do not need to analyze the user edits if it relates to content (e.g. specific dates)
        Output only a brief and concise summary of the edits. Do not infer motives. 
        Include relevant background context from the history — for example, when the change happened (if indicated) and how the text was modified. 
        Format: 'The user deleted XXX or added XXX to the {item\_type} when [context about the stage or progress of the allocation].' 
        Example: The user changed "Best regards," from the AI-generated outreach plan to "Thanks," in the email when initiating outreach to Jack regarding available dates.

Inputs:
history\_context: \{history\_context\}"
"ai\_text: \{ai\_text\}"
"user\_text: \{user\_text\}"
\end{promptbox}

\subsection{Edge Case Detection}
\label{subsec:edge-case-detection}

\hfill
\begin{promptbox}
  You are an expert seminar organizer assistant. 
        Analyze the email log, approved plans, time context, and reasoning summary. 

        You need to flag edge-case issues that may require user confirmation or special and urgent attention or the conditions are not covered by the coordination policies (e.g., last-minute availability change, medical or accessibility information, special requests like unusual conditions not covered by the policies).

        If an edge-case issue is found, output a concise 40-word message starting with "FLAG:" 
        that clearly explains the issue and ends with a direct question that requires user direction so the system can take action.

        Please note: If the user has already confirmed, is following the plan, or appears on the right track to solve the problem (e.g., they have already contacted speakers, have a plan in place, or are simply waiting for responses), then do not flag—output "Clear". 
        
        You need to output "Clear" for routine scheduling matters (e.g., unclaimed slots, unresponsive speakers, conflicting availability, or other common slot-assignment issues). These routine tasks can usually be resolved by following up or reaching out again. In comparison, edge-case issues require explicit user guidance and explanation to ensure the situation is addressed appropriately.

        \#\#\# FEWSHOT EXAMPLES \#\#\#

        Clear examples (no special action needed):
        Input: Many past emails… New Email: “Thanks for the invite, I’ll check my schedule and get back to you in a few days.”
        Output: Clear

        Input: Many past emails… One email from another speaker: “I want to come to Sep 1! Other slots may also work!” New Email: “I can only come Sep 1.”
        Output: Clear

        Flag examples (requires organizer decision):
        Input: Many past emails… New Email: “I’m on maternity leave this semester, not sure I can attend, maybe someone else from my lab could take the slot.”
        Output: FLAG: SpeakerName seems unavailable due to maternity leave and suggesting a lab substitute. Should I follow up with the proposed colleague or reassign the slot to another invited speaker?

        Input: Many past emails… New Email: “I’ll be abroad that entire week, is there any chance I can swap to a later slot in May?”
        Output: FLAG: SpeakerName's traveling during scheduled week requests swap to May. Should I reschedule them to May or retain current slot and reassign to another speaker?

        Input: Many past emails… New Email: “Hey! I am sorry!! I know my seminar is at the coming Friday! However, due to some family reason, I’ll only know the day before.”
        Output: FLAG: SpeakerName's uncertain due to family reason, with final notice only one day prior. Should I secure a backup speaker now or wait for their last-minute confirmation?

        Here are the policies: [Policies]
\end{promptbox}

\subsection{Simulated Respondents}
\label{subsec:simulated-respondents}

\hfill
\begin{promptbox}

            You are simulating a seminar speaker. Your persona: [Speaker Persona]. Name: [Speaker Name].

            The organizer's name is: [Organizer Name].

            The current simulation day is: [Day ID].
           
            You have received the following email thread (latest at the bottom):
[Email Log]

            Possible dates to choose from: [Speaker Availability]]

            As the speaker, you can:
            - Reply to the organizer's email, providing your availability, asking questions, or giving other relevant information.
            - If you have questions for the organizer, you can draft them in your reply and wait for a response.
            - You may choose to wait for more information before replying, if appropriate.
            - You can reply to multiple emails in the thread if you wish.
            - You may also choose to wait for a specific number of days before replying (e.g., "wait 2 days and reply on day 5"). If you choose to wait, specify the simulation day you intend to reply.

            Decide what you want to do next:
            - If you want to reply, write a detailed, persona-driven reply (possibly including questions for the organizer).
            - If you want to ask a question and wait for a response, draft your question(s) and indicate you are waiting. Do not ask questions that are too specific ('like equipment setup' etc, this is a scheduling task, they only care if you are available on the date.)
            - If you want to wait and not reply yet, explain why, and specify the simulation day you intend to reply (e.g., "wait\_until\_day": 5).

            Return your action as one of: 'reply', 'ask\_question', 'wait'.
            If you choose to wait, also return a "wait\_until\_day" field (integer simulation day to reply).

            Then, provide your reply email (if any) as a JSON object with keys: from, to, subject, body, timestamp. Remember, your subject should be just a short first-point-of-view summary of your reply (less than 10 words, like "I can make it on Sep 5", "I can't make it on Sep 5", "Sorry I need a few days to think." etc).

            Also, provide a short "RATIONALE" section (2-4 sentences) explaining your reasoning and thought process for this turn.

            Format your response as:
            ACTION: <reply|ask\_question|wait>
            WAIT\_UNTIL\_DAY: <integer or 'none'>
            EMAIL: <JSON object or 'none'>
            RATIONALE: <your reasoning>

            EMAIL should be like: {"from": "XX", "to": "XX", "subject": "XXX", "body": "XXXXX", "timestamp": "XX"}   dont output json ``` etc

\end{promptbox}

\subsection{Free-form QA Chat}
\label{subsec:freeform-qa-chat}

\hfill
\begin{promptbox}
You are an expert assistant. The following is an email log from a seminar simulation system. Answer the user's question based on the email log. If the answer is not directly available, say so, but try to be helpful and concise.

[EMAIL LOG]

[USER QUESTION]

\end{promptbox}

\subsection{Daily Summary}
\label{subsec:daily-summary}

\hfill
\begin{promptbox}

Today is simulation day [Day ID].
Here is the email log for today (may include previous days):

[Email Log]

Please provide a concise 40-word summary of the progress made today, including any key actions taken, responses received, and remaining issues. The summary should be suitable as a notification for users in a sidebar. Dont include today's date in the summary -- just summarize the progress.

\end{promptbox}

\section{Technical Evaluation Details}
\label{sec:technical-eval}

\subsection{Policy Selection}
Here are the prompts used in the technical evaluations for policy selection:
\subsubsection{Experiment prompts}
\paragraph{Speaker-only}
\hfill
\begin{promptbox}
You are a helpful assistant. Given the following context, select which of the following coordination policies should be applied. The chosen policies should be 100\% applied to the scenario! If you found that we didn't do anything yet from the reasoning, you should choose the 'we are just beginning' policy. Return a JSON list of the indices of the applicable policies (e.g., [1,2] or [1]). If none apply, return an empty list [].

Current context:
Pending Speakers:
\{pending\_speakers\_text\}

Return ONLY the JSON list of indices:
\end{promptbox}

\paragraph{TimeSlot-only}
\hfill
\begin{promptbox}
You are a helpful assistant. Given the following context, select which of the following coordination policies should be applied. The chosen policies should be 100\% applied to the scenario! If you found that we didn't do anything yet from the reasoning, you should choose the 'we are just beginning' policy. Return a JSON list of the indices of the applicable policies (e.g., [1,2] or [1]). If none apply, return an empty list [].

Current context:
Slot Status:
- Filled slots: \{filled\_slots\}
- Unfilled slots: \{unfilled\_slots\}

Return ONLY the JSON list of indices:
\end{promptbox}

\paragraph{TimeSlot + Speaker}
\hfill
\begin{promptbox}
You are a helpful assistant. Given the following context, select which of the following coordination policies should be applied. The chosen policies should be 100\% applied to the scenario! If you found that we didn't do anything yet from the reasoning, you should choose the 'we are just beginning' policy. Return a JSON list of the indices of the applicable policies (e.g., [1,2] or [1]). If none apply, return an empty list [].

Current context:
Slot Status:
- Filled slots: \{filled\_slots\}
- Unfilled slots: \{unfilled\_slots\}

Pending Speakers:
\{pending\_speakers\_text\}

Return ONLY the JSON list of indices:
\end{promptbox}

\paragraph{Summary-based}
\hfill
\begin{promptbox}
You are a helpful assistant. Given the following main reasoning result and context, select which of the following coordination policies should be applied. The chosen policies should be 100\% applied to the scenario! If you found that we didn't do anything yet from the reasoning, you should choose the 'we are just beginning' policy.

Return a JSON list of the indices of the applicable policies (e.g., [1,2] or [1]). If none apply, return an empty list [].

Main reasoning result:
\{progress\_summary\}

Coordination Policies:
\{policy\_list\}

Return ONLY the JSON list of indices:
\end{promptbox}

\subsection{Edge Case Detection}
\label{sec:technical-eval2}
\subsubsection{Test suite construction}
We employ GPT‑5 to generate a set of 10 edge-case (flaggable) test cases and 10 non-flagged (clear) test cases, with all examples validated for quality by the lead authors. To enhance linguistic and contextual diversity, each test case is further expanded into five stylistic variations, reflecting different tones and levels of formality. These variations are conditioned on five distinct speaker personas: a senior renowned faculty member, a junior faculty member, a senior PhD student, a junior PhD student, and a close peer researcher. In total, 100 test cases are constructed and used as our test suite for edge case detection.


\subsubsection{Prompts to generate linguistic variations and test data}
\hfill
\begin{promptbox}
You are a helpful assistant. You need to generate test cases for me to test my flag detection assistant. I am organizing a seminar series, inviting speakers. You will be acting as the simulated speakers, writing response emails to my invitation email.
You will be given cases where the system should mark as clear and that are not edge cases. For each case, you should generate exactly five different variations of responses in the role of the following five speaker personas:
1) a famous distinguished professor who is always busy
2) a junior faculty who is trying to establish herself in the field
3) a senior phd student who is able to graduate and is looking for a job
4) a junior phd student who is just starting his phd journey
5) a close friend of mine who is also a phd student researcher.
The following are the cases where the system should mark as clear (i.e. labeled as "0"):
Please generate the test cases in the following JSON format: """
\{
"case\_description": \[
\{
"persona": "persona\_description",
"response": "response",
"flag": "0"
\}
\]
\}
The case\_description is the title of each scenario shown above. Please generate exactly five variations of the response for each case

\end{promptbox}

\subsubsection{Experiment prompts}
\paragraph{zero-shot}
\hfill
\begin{promptbox}
You are an expert seminar organizer assistant.
Analyze this response from a speaker invitation. 

Response: {response}

You need to flag edge-case issues that may require user confirmation or special and urgent attention

If an edge-case issue is found, simply output "FLAG", otherwise output "Clear".

Your Output:
\end{promptbox}

\paragraph{few\_shot without policy}
\hfill
\begin{promptbox}
You are an expert seminar organizer assistant. 
        Analyze the email log, approved plans, time context, and reasoning summary. 

        You need to flag edge-case issues that may require user confirmation or special and urgent attention or the conditions are not covered by the coordination policies (e.g., last-minute availability change, medical or accessibility information, special requests like unusual conditions not covered by the policies).

        If an edge-case issue is found, output a concise 40-word message starting with "FLAG:" 
        that clearly explains the issue and ends with a direct question that requires user direction so the system can take action.

        Please note: If the user has already confirmed, is following the plan, or appears on the right track to solve the problem (e.g., they have already contacted speakers, have a plan in place, or are simply waiting for responses), then do not flag—output "Clear". 
        
        You need to output "Clear" for routine scheduling matters (e.g., unclaimed slots, unresponsive speakers, conflicting availability, or other common slot-assignment issues). These routine tasks can usually be resolved by following up or reaching out again. In comparison, edge-case issues require explicit user guidance and explanation to ensure the situation is addressed appropriately.

        \#\#\# FEWSHOT EXAMPLES \#\#\#

        Clear examples (no special action needed):
        Input: Many past emails… New Email: “Thanks for the invite, I’ll check my schedule and get back to you in a few days.”
        Output: Clear

        Input: Many past emails… One email from another speaker: “I want to come to Sep 1! Other slots may also work!” New Email: “I can only come Sep 1.”
        Output: Clear

        Flag examples (requires organizer decision):
        Input: Many past emails… New Email: “I’m on maternity leave this semester, not sure I can attend, maybe someone else from my lab could take the slot.”
        Output: FLAG

        Input: Many past emails… New Email: “I’ll be abroad that entire week, is there any chance I can swap to a later slot in May?”
        Output: FLAG

        Input: Many past emails… New Email: “Hey! I am sorry!! I know my seminar is at the coming Friday! However, due to some family reason, I’ll only know the day before.”
        Output: FLAG

        Response: \{response\}

        Your Output:
\end{promptbox}

\paragraph{few\_shot with policy}
\hfill
\begin{promptbox}
 You are an expert seminar organizer assistant. 
        Analyze the email log, approved plans, time context, and reasoning summary. 

        You need to flag edge-case issues that may require user confirmation or special and urgent attention or the conditions are not covered by the coordination policies (e.g., last-minute availability change, medical or accessibility information, special requests like unusual conditions not covered by the policies).

        If an edge-case issue is found, output a concise 40-word message starting with "FLAG:" 
        that clearly explains the issue and ends with a direct question that requires user direction so the system can take action.

        Please note: If the user has already confirmed, is following the plan, or appears on the right track to solve the problem (e.g., they have already contacted speakers, have a plan in place, or are simply waiting for responses), then do not flag—output "Clear". 
        
        You need to output "Clear" for routine scheduling matters (e.g., unclaimed slots, unresponsive speakers, conflicting availability, or other common slot-assignment issues). These routine tasks can usually be resolved by following up or reaching out again. In comparison, edge-case issues require explicit user guidance and explanation to ensure the situation is addressed appropriately.

        \#\#\# FEWSHOT EXAMPLES \#\#\#

        Clear examples (no special action needed):
        Input: Many past emails… New Email: “Thanks for the invite, I’ll check my schedule and get back to you in a few days.”
        Output: Clear

        Input: Many past emails… One email from another speaker: “I want to come to Sep 1! Other slots may also work!” New Email: “I can only come Sep 1.”
        Output: Clear

        Flag examples (requires organizer decision):
        Input: Many past emails… New Email: “I’m on maternity leave this semester, not sure I can attend, maybe someone else from my lab could take the slot.”
        Output: FLAG

        Input: Many past emails… New Email: “I’ll be abroad that entire week, is there any chance I can swap to a later slot in May?”
        Output: FLAG

        Input: Many past emails… New Email: “Hey! I am sorry!! I know my seminar is at the coming Friday! However, due to some family reason, I’ll only know the day before.”
        Output: FLAG

        Here are the policies:
        -   "Don't be shy to ask for more availability if they don't give you many. They will probably reply more if you beg."
        -   "Some people don’t read or respond to emails right away because they’re busy. You should wait for their response, but it’s okay to follow up every two workday days if you haven’t heard back.",
        -   "If someone hasn’t responded to the email after three reach-outs, then don’t contact them again.",
        -   "Ambiguity is okay. People can typically do the days they are vague about — you can prioritize asking for more availability and just confirm the ambiguity right before assigning them at the end.",
        -   "Some speakers—especially faculty—are more desirable to schedule. We should prioritize their availability by assigning slots to faculty over students and asking students to be more flexible or to provide additional dates to accommodate them.",
        -   "If we are just beginning and have not yet contacted any speakers for availability, then ask each speaker for their availability across all slots."

        Response: {response}

        Your Output:
\end{promptbox}

\onecolumn
\newpage
\section{System Screenshots}

\begin{figure}[]
\centering
\begin{subfigure}{.5\linewidth
}
  \centering
  \includegraphics[width=\linewidth]{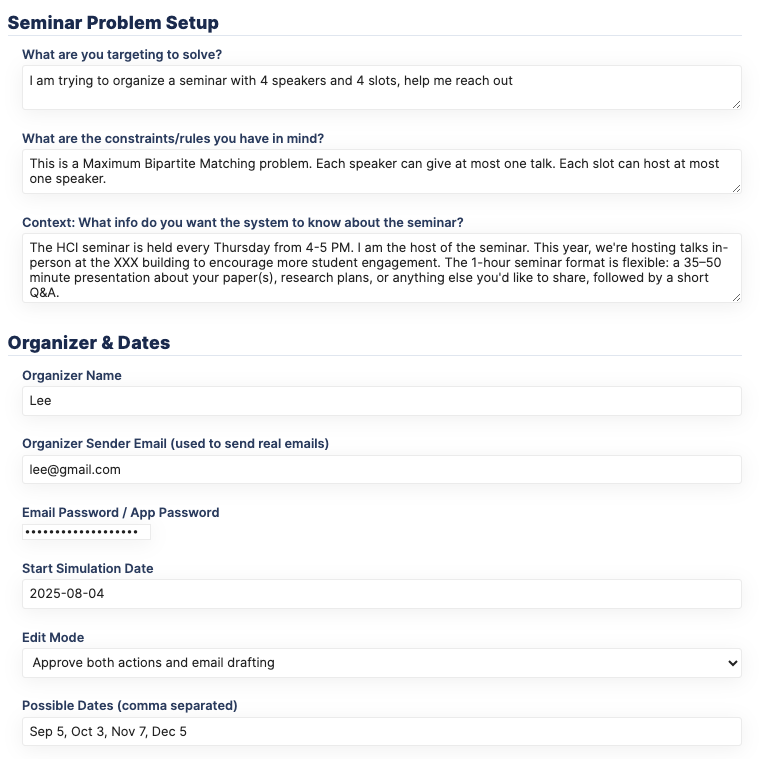}
  \label{fig:sub1}
\end{subfigure}%
\hfill
\begin{subfigure}{.5\linewidth}
  \centering
  \includegraphics[width=\linewidth]{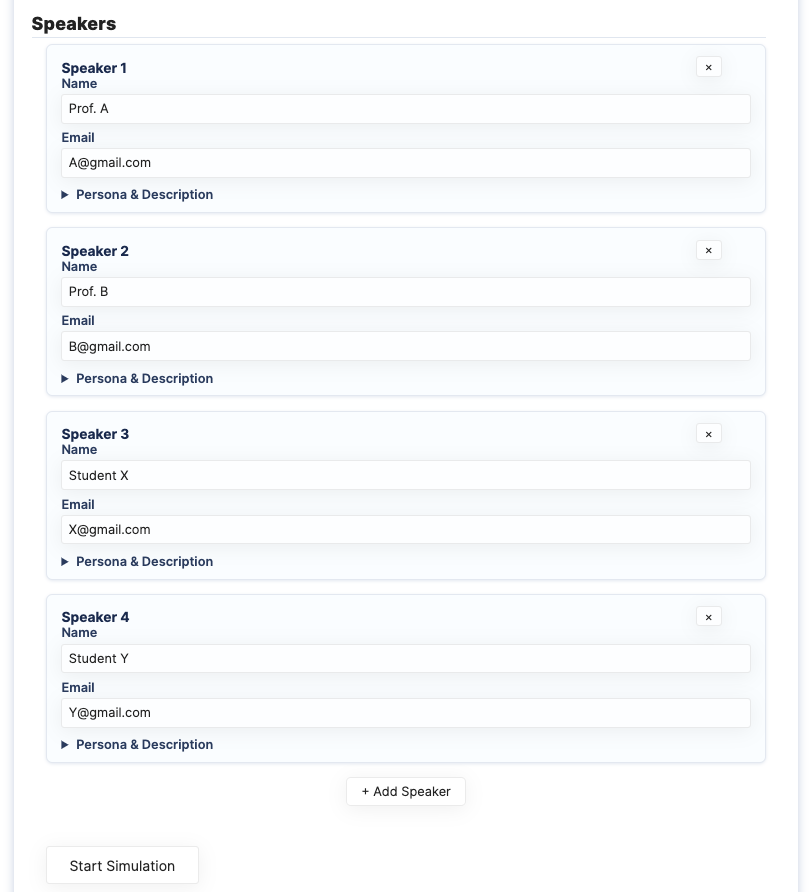}
  \label{fig:sub2}
\end{subfigure}
\caption{The landing page to specify the user goals, seminar details, and the persona info for the speakers.}
\label{fig:landing_page}
\end{figure}

\begin{figure}[!t]
\centering
\begin{subfigure}{.48\linewidth}
  \centering
  \includegraphics[width=\linewidth]{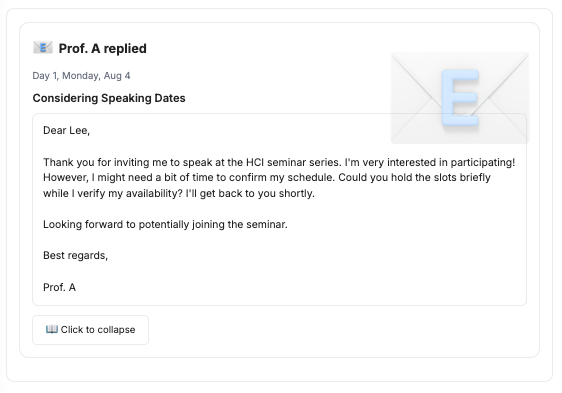}
  \caption{Email sent by the simulated Prof. A.}
  \label{fig:email_receive_profA}
\end{subfigure}%
\hfill
\begin{subfigure}{.48\linewidth}
  \centering
  \includegraphics[width=\linewidth]{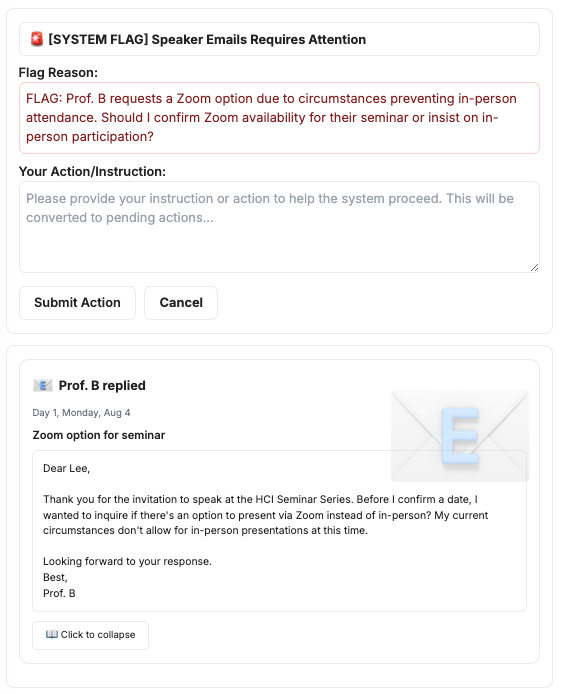}
  \caption{Edge case detection given response from Prof. B asking about Zoom options.}
  \label{fig:flag_email}
\end{subfigure}
\caption{The example email sent by the simulated respondents and the system message when detecting edge case.}
\label{fig:email_receive}
\end{figure}



\newpage

\end{document}